\pdfoutput=1 
\documentclass[preprint,12pt]{elsarticle}

\usepackage[english]{babel}
\usepackage{blindtext}
\usepackage[nolist]{acronym}
\usepackage{amssymb}
\usepackage{amsmath}
\usepackage[font=small,skip=3pt]{caption}
\usepackage[font=small,skip=6pt]{subcaption}
\usepackage{tikz}
\usepackage{pgfplots}
\pgfplotsset{compat=1.18} % sugention from a compiler warning -> removed warning
\usepackage{pgfplots,pgfplotstable}

\newcommand\maxnorm[1]{\|#1\|^{\text{max}}}

\newcommand\Bigmaxnorm[1]{\Big\|#1\Big\|^{\text{max}}}

\journal{arXiv}

\begin{document}

\begin{frontmatter}



\title{A variable speed of sound formulation for weakly compressible smoothed particle hydrodynamics}


\author[inst1]{Fabian Thiery}

\affiliation[inst1]{organization={Munich Institute of Integrated Materials, Energy and Process Engineering (MEP), Technical University of Munich (TUM)},%Department and Organization
            addressline={Lichtenbergstr. 4a}, 
            city={Garching},
            postcode={85748}, 
            % state={State One},
            country={Germany}}

\author[inst1,inst2]{Nikolaus A. Adams}

\affiliation[inst2]{organization={Chair of Aerodynamics and Fluid Mechanics, Technical University of Munich (TUM)},%Department and Organization
            addressline={Boltzmannstr. 15}, 
            city={Garching},
            postcode={85748}, 
            % state={State Two},
            country={Germany}}
            
\author[inst1]{Stefan Adami}

\begin{abstract}
We present a \ac{wcsph} formulation with a temporally variable speed of sound.
The benefits of a time-varying sound speed formulation and the weaknesses of a constant sound speed formulation are worked out. 
It is shown how a variable sound speed can improve the performance, accuracy, and applicability of the \ac{wcsph} method.
In our novel \ac{ucsph} method, the required artificial speed of sound is calculated at each time step based on the current flow field. 
The method's robustness, performance, and accuracy are demonstrated with three test cases: a Taylor-Green vortex flow, a falling droplet impact, and a dam break.
For all showcases, we observe at least similar accuracy as computed with \ac{wcsph} at strongly improved computational performance.  
\end{abstract}

\acresetall


% \begin{highlights}
% \end{highlights}


\end{frontmatter}


% INCLUDE THE ACRONYMS
\begin{acronym}

% SPH Typs
\acro{sph}[SPH]{Smoothed Particle Hydrodynamics}
\acro{isph}[ISPH]{Truly Incompressible SPH}
\acro{wcsph}[WCSPH]{Weakly Compressible SPH}
\acro{ucsph}[UCSPH]{Uniform Compressible SPH}
\acro{pcisph}[PCISPH]{Predictive Corrective Incompressible SPH}
\acro{dfsph}[DFSPH]{Divergence-Free SPH}
\acro{iisph}[IISPH]{Implicit incompressible SPH}
\acro{gpu}[GPU]{Graphics processing unit}

% Other
\acro{eos}[EOS]{equation of state}

\end{acronym}

% INCLUDE THE INTRODUCTION
\newpage
\section{Introduction}
\label{sec:introduction}
%
%% SPH in general:
\ac{sph} is a fully Lagrangian, mesh-free method widely used in computational fluid dynamics. 
Initially, the method was developed for astrophysical problems by \citet{Gingold1977} and \citet{Lucy1977} in 1977.
In \ac{sph}, the domain is discretized with a finite number of particles. 
Each particle represents a fluid element with a specific mass.
The method relies on a so-called smoothing kernel.
The smoothing kernel is used to calculate a particle's density, acceleration, and other quantities based on its neighbors.
Each particle is shifted in time along its computed trajectory. 
Since 1977, the method has been further developed and applied to 
many kinds of fluid dynamics problems, such as multi-phase flows \cite{Monaghan1994MultiPhase, Colagrossi2003, Chen2015}, free-surface flows \cite{Monaghan1994FreeSurface, Violeau2016}, porous media flows \cite{Holmes2011, Holmes2021, Osorno2021}, surface tension driven flows \cite{Morris2000, Hu2006, Blank2023}, and flows with species transport \cite{Adami2010Surfactant, Kim2023}. 
Most engineering flow problems simulated with SPH are considered to be incompressible.
Generally, two different approaches exist in \ac{sph} to handle such flows: \ac{isph} and \ac{wcsph}.

%
%% weakly compressible:
For the weakly compressible modeling, a fluid is considered compressible, but its compressibility is limited to the artificial compressibility $\delta_\rho$.  
A commonly used constraint for weakly compressible fluids is that artificial compressibility $\delta_\rho$ has to be smaller than 1\% for the occurring forces. 
This relates to a Mach number $Ma = u / c_s$ smaller than  $0.1$.
Due to the CFL-time step criterion, weakly compressible solvers have a much smaller time step than truly incompressible solvers.
On the other hand, due to the local and fully explicit formulation of a weakly compressible solver, the computational cost of one time step is much lower.

%
%% truly incompressible: 
In truly incompressible model approaches, the density of the fluid is considered constant, and the velocity field is required to be divergence-free. 
%
% ISPH:
This is typically enforced by a pressure projection method \cite{Cummins1999, Lind2012}. 
The method enables large time steps but has a high computational cost per time step due to the elliptic nature of the problem.
Recently, the popularity of novel iterative incompressible solvers increased in the field of computer graphics and partially also in computational fluid dynamics \cite{Solenthaler2009, Ihmsen2014, Bender2015, Peng2019}.
These types of solvers ensure incompressibility by an iterative process.
The pressure is computed with an equation similar to the \ac{eos} used in \ac{wcsph}.
However, unlike \ac{wcsph}, the used pressure equation varies temporally and spatially.

%
% Comparison of ISPH and WCSPH: 
Both methods, \ac{wcsph} and \ac{ucsph}, are frequently used, and each has pros and cons.
For a comparison of both methods, see \cite{Chen2013,Lee2008}.
This work focuses on the weakly compressible approach for its ease of implementation and its parallel algorithm well suited for \ac{gpu} implementations. 
In the following, we point out some weaknesses of the classical weakly compressible model approach and present our adjustments to overcome them. 

%
%% EOS:
In weakly compressible fluids, the pressure is a function of the density.
The relation of the two quantities is given by a so-called \ac{eos} \cite{Cole1948, Morris1997}.
The stiffness of the \ac{eos} depends on the maximal pressure and velocity appearing in a simulation - the higher the pressure and velocity, the stiffer the equation, and the smaller the time step for a stable time integration.
In classical \ac{wcsph}, the \ac{eos} stiffness is constant throughout the simulation, or in other words, the speed of sound is constant. 
%
%% Variable time stepping
When using adaptive-time stepping schemes, like those from \citet{Monaghan1999} and \citet{Crespo2015}, this constant speed of sound introduces an upper limit for the adaptive time steps $\Delta t_{\text{max}} = 0.25 h/c_s$. 
For unsteady flow configurations with highly changing pressure and velocity magnitudes, at some point, this upper time step limit can be unnecessarily small. 
% 
%% Teaser for our work:
To the best of our knowledge, all published \ac{wcsph} solvers use a constant speed of sound per phase.
In this paper, we demonstrate that using a variable speed of sound can significantly decrease the computational costs.
We also show that the stiffness of the \ac{eos} influences the method's accuracy. 
Furthermore, we tackle an issue in classical \ac{wcsph} solvers, which is the definition of the constant speed of sound. 
This a-priori definition is non-trivial for complex cases.    
Our novel \ac{ucsph} method is introduced in the following sections.

% INCLUDE THE SECTION GOVERNING EQUATIONS
% \newpage
\section{Governing Equations}
\label{sec:governing_equations}
The governing equations for the motion of an isothermal fluid in a Lagrangian frame are given by the mass conservation law
\begin{equation}
    \label{eq:mass_conservation}
    \frac{d\rho}{dt} = -\rho \nabla \cdot \vec{v}
    \text{,}
\end{equation}
and the momentum conservation law
\begin{equation}
    \label{eq:momentum_conservation}
    \rho\frac{d\vec{v}}{dt} = -\nabla p + \vec{F}^{(\nu)} + \rho \vec{g}
    \text{,}
\end{equation}
with $\rho, t, \vec{v}, p, \vec{F}^{(\nu)}, g$ denoting density, time, velocity, pressure, viscous force, and body force, respectively.
When assuming incompressibility, the viscous force can be written as
\begin{equation}
    \label{eq:viscous_force}
    \vec{F}^{(\nu)} = \eta \nabla^2 \vec{v},
\end{equation}
where $\eta$ is the dynamic viscosity of the fluid.
By following the weakly compressible approach \cite{Cole1948, Morris1997},
an \ac{eos} is used to compute the pressure as a function of the density
\begin{equation}
    \label{eq:eos_complete}
    p(\rho) = p_0\Big[\Big(\frac{\rho}{\rho_0}\Big)^\gamma - 1\Big]
    \textnormal{.}
\end{equation}
The variable $\rho_0$ denotes the reference density of the fluid.
The reference pressure $p_0$ and the exponent $\gamma$ are used to define the problem-specific stiffness of the \ac{eos}.

% INCLUDE THE SECTION NUMERICAL METHOD
\section{Numerical Method}
\label{sec:numerical_method}
This section presents the classical \ac{wcsph} method and the changes leading to our novel \ac{ucsph} method. 
The conservation of mass in Sec. \ref{sec:conservation_of_mass} and the conservation of momentum in Sec. \ref{sec:conservation_of_momentum} are for \ac{wcsph} and \ac{ucsph} equivalent. 
In Sec. \ref{sec:equation_of_state} the classical \ac{eos} and the stability limits of \ac{wcsph} are shown. 
Our novel \ac{ucsph} method is presented in Section \ref{sec:equation_of_state_with_variable_speed_of_sound}.
%
%% Mass conservation: 
\subsection{Conservation of Mass}
\label{sec:conservation_of_mass}
In \ac{sph}, there are two ways to compute the density.
First, the density evolution based on the continuity equation from Eq. \eqref{eq:mass_conservation}, and second, the density summation based on a summation over neighboring particles.
The presented \ac{sph} solver relies on a unified formulation for the density summation approach
\begin{equation}
    \rho_a = \text{max}\Big(\rho_0, \sum_b m_b W_{ab} + \text{max}\Big(1-\sum_b V_b W_{ab}, 0\Big)\rho_0\Big)
\end{equation}
to handle free-surface and bulk flows; see \citet{Hahn2018} or \citet{Zhang2022}.
Additionally, the density evolution 
\begin{equation}
    \label{eq:time_derivative_density_single_phase}
    \frac{d\rho}{dt} = 
    \rho_a
    \sum_b 
        \vec{v}_{ab} 
        V_b 
        \frac{\partial W_{ab}}{\partial r_{ab}} 
        \vec{e}_{ab} 
\end{equation}
is performed inside the used Verlet time stepping scheme \cite{Verlet1967} to get a more accurate density at the Verlet stages.

%
%% Momentum conservation: 
\subsection{Conservation of Momentum}
\label{sec:conservation_of_momentum}
The momentum conservation law from Eq. \eqref{eq:momentum_conservation} is divided into three terms.
%
%% Pressure acceleration
First, the acceleration due to the pressure gradient, given by
\begin{equation}
    \frac{d\vec{v}^{(p)}_a}{dt} =
    -
    \sum_b 
        m_b
        \Big(\frac{p_a}{\rho_a^2}+\frac{p_b}{\rho_b^2}\Big)
        \frac{\partial W_{ab}}{\partial r_{ab}}
        \vec{e}_{ab} 
        \textnormal{.}
    \label{eq:pressure_acceleration}
\end{equation}
%
%% Viscous acceleration 
Secondly, the viscous acceleration via 
\begin{equation}
    \label{eq:viscous_acceleration}
    \frac{d\vec{v}^{(\nu)}_a}{dt} = 
    \frac{\eta}{m_a}
    \sum_b
        (V_a^2 + V_b^2)
        \frac{\vec{v}_{ab}}{r_{ab}}
        \frac{\partial W_{ab}}{\partial r_{ab}}
\end{equation}
for incompressible fluids \cite{Hu2006}.
%
%% Body-force
The third term in Eq. \eqref{eq:momentum_conservation} is the body force $g$, a constant acceleration, e.g., gravity, acting on each particle in the domain.

%
%% Equation of state:
\subsection{Equation of State}
\label{sec:equation_of_state}
In a weakly compressible formulation, the pressure is a function of the density.
The relation of the two quantities is given by the \ac{eos} in Eq. \eqref{eq:eos_complete}.
With the  reference pressure
\begin{equation}
    \label{eq:reference_pressure}
    p_0 = \frac{c_s^2 \rho_0}{\gamma}
    \text{,}
\end{equation}
the \ac{eos} results in the widely used Cole equation \cite{Cole1948}
\begin{equation}
    \label{eq:eos_complete_cole}
    p(\rho) = \frac{c_s^2 \rho_0}{\gamma}\Big[\Big(\frac{\rho}{\rho_0}\Big)^\gamma - 1\Big]
    \textnormal{.}
\end{equation}
Therefore, a user has to determine a suitable artificial speed of sound for the problem of interest. 
The requirements for the artificial speed of sound 
\begin{equation}
    \label{eq:speed_of_sound_criterion}
    c_s = \text{\text{max}}\Bigg(10 v_{\text{max}}, \sqrt{\frac{p_{\text{max}}\gamma}{\rho_0[(1+\delta_\rho)^\gamma-1]}}\Bigg)
\end{equation}
are derived from the admissible compressibility $\delta_\rho = (\rho / \rho_0 -1)$, the maximal expected velocity $v_{\text{max}}$, and the maximal expected pressure in the flow $p_{\text{max}}$. 
The used sound speed criteria, based on $v_{\text{max}}$ and $p_{\text{max}}$ is a generalized form of the well known and widely used formulation from \citet{Morris1997}.

Vanilla \ac{wcsph} uses a minimal time step for the time integration based on a stability analysis of the right hand side terms, see \cite{Morris1997}. 
In this work, we use an adaptive time stepping scheme considering the maximal velocity and maximal acceleration via
\begin{equation}
    \label{eq:time_step_criterion_cfl}
    \Delta t_{\text{cfl}} = 0.25 \frac{h}{\maxnorm{\vec{v}_a(t)}+c_s}
    \text{,}
\end{equation}
\begin{equation}
    \label{eq:time_step_criterion_acceleration}
    \Delta t_{\text{acc}} = 0.25 \sqrt{\frac{h}{\maxnorm{\frac{d\vec{v}}{dt}\big|_a(t)}}}
    \text{,}
\end{equation}
\begin{equation}
    \label{eq:time_step_criterion}
    \Delta t = \text{min}(\Delta t_{\text{cfl}}, \Delta t_{\text{acc}})
    \text{,}
\end{equation}
similar to the work of \citet{Monaghan1999} and \citet{Crespo2015}.
For many flow problems, the CFL criterion defines the minimal time step. 
Note, given that $c_s \geq 10 v_{\text{max}}$, the performance improvement using this time-step adaption is marginal. 
%
%% Equation of state with variable speed of sound:
\section{Uniform Compressible SPH}
\label{sec:equation_of_state_with_variable_speed_of_sound}
%
%% Motivation: 
If we recap the constraints for the artificial speed of sound in \ac{wcsph}, we observe that it only depends on the highest pressure and velocity. 
The artificial speed of sound has to be chosen such that the weakly compressible assumptions are even valid for these worst-case time steps.
For all the other time steps where the maximal occurring pressures and velocities may be much smaller, the speed of sound is unnecessarily high, leading to unnecessary low time steps. 
To overcome these limitations, we introduce a novel method where the speed of sound is adapted to the current flow field, which results in a time-dependent speed of sound.
The time-dependent speed of sound influences the compressibility of the fluid, such that at each time step, the maximal compression of the fluid is close to 1\%. 
For that reason, we name our novel form of the \ac{wcsph} method \acf{ucsph}. 
\subsection{Basic Algorithm}
The basic algorithm relies on a frequently updated maximal velocity and maximal pressure in the speed of sound formula. 
The maximal velocity and the maximal pressure in the flow field
\begin{equation}
    \label{eq:max_velocity_prediction}
    v^{(n)}_{\text{max}} = 
    \Bigmaxnorm{
    \vec{v}^{(n)}_a
    }
    \text{,}
\end{equation}
\begin{equation}
    \label{eq:max_pressure_prediction}
    p^{(n)}_{\text{max}} = 
    \Bigmaxnorm{
    p^{(n)}_a 
    }
\end{equation}
are computed at the end of each time step.
The speed of sound is then updated by
\begin{equation}
    c_s^{(n+1)} = 
    \max 
    \Bigg(
    10 v^{(n)}_{\text{max}},
    \sqrt{\frac{p^{(n)}_{\text{max}}\gamma}{\rho_0[(1+\delta_\rho)^\gamma-1]}}
    \Bigg)
    \text{.}
    \label{eq:speed_of_sound_prediction_naive}
\end{equation}
Based on the updated speed of sound, the time step length of the next time integration can be computed. 
Here, a similar adaptive time stepping scheme is used as in \ac{wcsph}, given in Eq. \eqref{eq:time_step_criterion_cfl} - Eq. \eqref{eq:time_step_criterion}.
The only difference is the time-dependent speed of sound. 
The result is a significant increase in the degree of adaptivity of the time step.  
With the \ac{ucsph} method, the adaptive time step is no longer limited by the constant speed of sound in the CFL-time step.
With the variable speed of sound, the CFL-time step can become orders of magnitudes higher than with a constant speed of sound.
Clearly, this influences the performance of the method significantly. 

Since the presented algorithm automatically adjusts the speed of sound, there is no need for a maximal pressure and velocity value before the simulation starts, as usual in \ac{wcsph}.  
This is a large benefit of the novel method and relieves the efficient simulation of complex cases where it is hard to predict the maximal pressure and velocity in advance. 
Nevertheless, an initial speed of sound is still needed. 
Therefore, the user has to provide the initial velocity and pressure of the simulation.

The presented basic algorithm is not robust for violent flows and could introduce significant instabilities. 
The reason, therefore, is the pressure formulation in the weakly compressible modeling. 
As known, the pressure is a scaled density, where the speed of sound is one scaling factor. 
When changing the speed of sound from one step to the other, while the density stays constant, the pressure is artificially changed.
This artificial pressure change can introduce spurious particle movements and instabilities. 
However, when the order of the artificial pressure change is small, the solution of interest is not significantly influenced.
To minimize the artificial pressure change, various approaches are conceivable. 
Firstly, using a pressure formulation where the pressure is not directly influenced by changing the speed of sound, similar to the EDAC schemes \cite{Ramachandran2019}.
Secondly, an additional shifting scheme is conceivable, which influences the density in such a way that after the speed of sound adjustment, the same pressure is present as before the adjustment. 
The third approach, which is studied in the present paper, is to adjust the speed of sound in such a way that the artificial pressure changes are small and do not influence the overall solution. 
The benefit of the chosen approach is that no shifting scheme and additional pressure formulations are needed. 
On the other hand, the algorithm to adjust the speed of sound increases in complexity.
In the following, multiple additions to the basic algorithm are introduced to obtain a robust variable speed of sound formulation with minimal influence on the pressure field.  

\subsection{Variable Speed of Sound}
Given the compressible nature of the method, \ac{wcsph} is prone to pressure oscillations for any discontinuous change in the flow field.
Such numerical high-frequency effects should not be considered for the dynamic speed of sound adjustments. 
Therefore, we only use the maximal sound speed from a the previous time interval $t_{\text{history}}$ to adjust the \ac{eos} smoothly. 
This time interval $t_{\text{history}}$ should be large enough to filter out purely numerical fluctuations yet small enough to identify the physical flow evolution. 
To ensure that, the user has to provide a reference length $\tilde{L}$, which should be the maximal length where pressure waves could be reflected inside the domain.
For example, in a thin hydrostatic tank, the reference length would be the depth of the tank.
This length determines the frequency of the pressure oscillations, and accordingly, the time interval
\begin{equation}
    \Delta t_{\text{history}} 
    =
    \frac{4\tilde{L}}{\bar{c}_s}
    \text{.}
\end{equation}
The speed of sound prediction for the next time step is then bounded by 
\begin{equation}
    c_s^{(n+1)} = \max     \big(c_s^{(n+1)}, \bar{c}_s \big)
    \text{.}
\end{equation}

In order to limit the artificial pressure changes due to the changing speed of sound, we introduce an upper limit for the sound speed change per time step $|\Delta c_s| = |c_s^{(n+1)} - c_s^{(n)}|$ given by
\begin{equation}
    |\Delta c_s| < \epsilon c_s^{(n)}
    \text{.}
\end{equation}
In this work, $\epsilon$ is set to 1\%. 
The sound speed change per time step can also be written as 
\begin{equation}
    |\Delta c_s| = \frac{\partial c}{\partial t} \Delta t
    \text{,}
\end{equation}
where 
\begin{equation}
    \frac{\partial c}{\partial t} = \frac{c_s^{(n+1)}-c_s^{(n)}}{\Delta t^{(n)}}
\end{equation}
is the temporal derivative of the sound speed. 
When rearranging the equations, an additional time step criterion can be derived
\begin{equation}
    \Delta t_{\epsilon} 
    = \frac{|\Delta c_s|}{\frac{\partial c_s}{\partial t}}
    = \frac{\epsilon c_s^{(n)}}{\frac{\partial c_s}{\partial t}} 
    = \frac{\epsilon c_s^{(n)} \Delta t^{(n)}}{c_s^{(n+1)}-c_s^{(n)}} 
    \text{.}
\end{equation}
The presented time step criterion ensures that the speed of sound never changes more than 1\% within a time step. 
We have observed that this additional time step criterion enhances stability and smoothness of the results significantly. 
Due to the additional time step criterion $\Delta t_\epsilon$, a permanent speed of sound adjustment could lead to a performance decrease. 
Therefore, if all of the following conditions are true   
\begin{equation}
    \begin{aligned}
    c_s^{(n+1)} < c_s^{(n)} < 1.25 c_s^{(n+1)}
    \label{eq:sound_speed_adjustment_criteria_1}
    \end{aligned}
\end{equation}
\begin{equation}
    \begin{aligned}
    0.8\% < \delta_{\rho} < 1.0\%
    \label{eq:sound_speed_adjustment_criteria_2}
    \end{aligned}
\end{equation}
we do not adjust the speed of sound.
Once a single criterion is false, the speed of sound is adjusted. 
It must be mentioned that the method works as well with other limits or even without the criteria.
Nevertheless, using these criteria leads to a significantly better performance in all our considered cases. 

% INCLUDE THE SECTION RESULTS
\section{Results}
\label{sec:results}
In this section, we demonstrate the accuracy, robustness, and performance of our presented \ac{ucsph} method in comparison to the classical \ac{wcsph} approach. 
We simulate all cases with the \ac{ucsph} and the \ac{wcsph} method to compare the solutions. 
Where available, analytical solutions are used for the validation. 
The first example is the Taylor-Green vortex flow. 
Due to the decaying nature of the case, it is ideal for showing the advantages regarding the performance and accuracy of the \ac{ucsph} method.
The following example is the falling droplet case, for which is hard to define the proper speed of sound a-priori. 
Finally, we present the well-known dam break case to confirm the robustness of \ac{ucsph} for violent impact scenarios.
%
%% Taylor-Green vortex
\subsection{Taylor-Green Vortex}
\label{sec:taylor_green_vortex}
The Taylor-Green vortex is a well-known validation case to demonstrate the accuracy of a numerical method. The case has an analytical solution for the two-dimensional incompressible Navies-Stokes Equations, given by
\begin{equation}
    \begin{aligned}
        u(x,y,t) &= -Ue^{bt} cos(2 \pi x) sin(2 \pi y)\\
        v(x,y,t) &= Ue^{bt} sin(2 \pi x) cos(2 \pi y)\\
        p(x,y,t) &= \frac{\rho}{4} (cos(4 \pi x) + sin(4 \pi y)) e^{(bt)^2}
    \label{eq:tgv_analytical}
    \end{aligned}
\end{equation}
with the exponential factor $b = -8\pi^2/Re$ and the maximal velocity $U$.
\begin{figure}[ht]
    \centering
    \includegraphics[width=0.5\columnwidth]{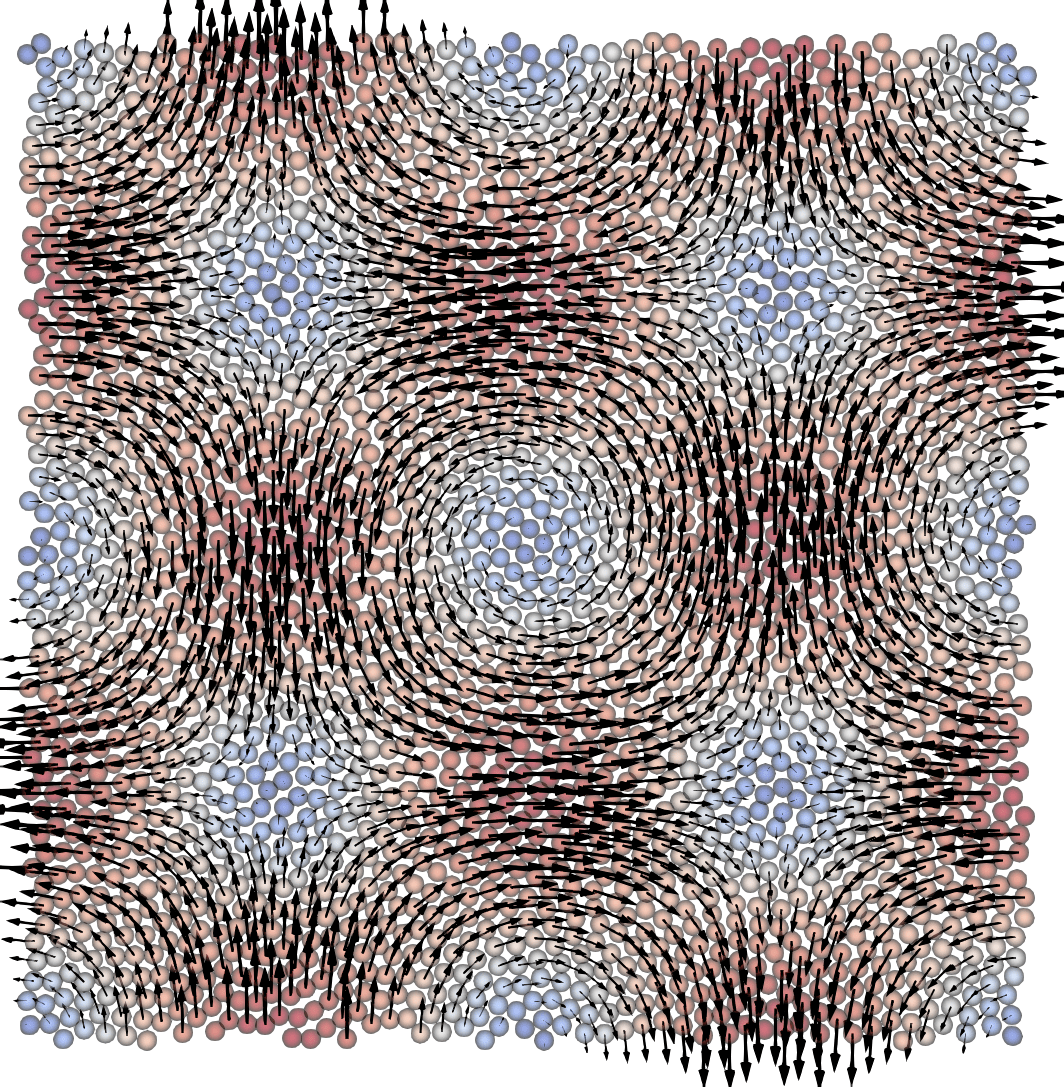}
    \caption{The initial velocity field of the Taylor-Green vortex.}
    \label{fig:Velocity_field_at_t=0}
\end{figure}
The domain has a side length of $L=1$ in the $x$- and $y$-direction with periodic boundaries.
The initial velocity field at $t = 0$ is set according to Eq. \eqref{eq:tgv_analytical} with $U=1$. 
The kinematic viscosity of the fluid is set to $\nu = 0.01$  such that $Re = U L / \nu = 100$. 
The initial particle spacing is $\Delta x = 0.02$, resulting in 2500 particles.
Two common initialization strategies for the Taylor-Green vortex exist in \ac{sph}. 
In the first one, the particles are placed on a Cartesian grid. 
Alternatively, a relaxed particle configuration can be employed.
This is often achieved from a pre-simulation with strong kinetic energy decay. 
Since starting from a Cartesian particle distribution introduces spurious velocities due to particle rearrangements at the beginning of the simulation \cite{Adami2013}, we use the second initialization option. 
For both methods, \ac{wcsph} and \ac{ucsph}, the speed of sound is initialized as $c_s = 10$.

\begin{figure}[ht]
    \centering
    \input{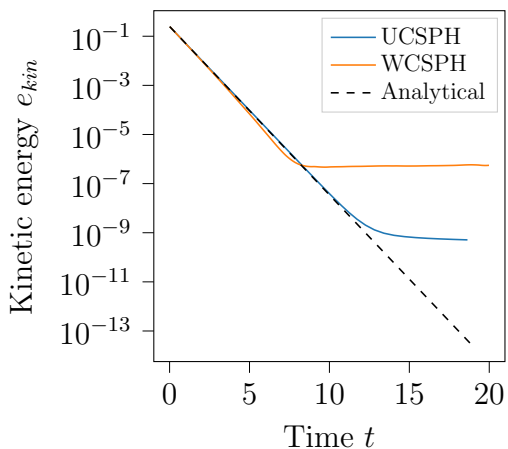}
    \caption{The kinetic energy decay of the Taylor-Green vortex at Re=100. }
    \label{fig:tgv_ekin}
\end{figure}

The initial velocity field and the vortex structure of the Taylor-Green vortex are visualized in Fig. \ref{fig:Velocity_field_at_t=0}.
Due to viscous damping, the vortex rotation decreases over time, resulting in a kinetic energy decay.
The decay is given by 
\begin{equation}
    e_{kin} = 0.5 U e^{2bt}
    \label{eq:analytical_kinetic_energy_decay}
\end{equation}
for a laminar flow field.
The numerically predicted kinetic energy decay can be compared against the analytical solution from Eq. \eqref{eq:analytical_kinetic_energy_decay} as validation.
Fig. \ref{fig:tgv_ekin} plots the analytical decay against the predicted decay from the \ac{ucsph} and the \ac{wcsph} method. 
Theoretically, the kinetic energy approaches zero for $\lim_{t\to\infty}$.
Due to numerical errors, the simulated solution diverges from the analytical at some point.
In Fig. \ref{fig:tgv_ekin}, it can be seen that the solution from the \ac{wcsph} follows the analytical solution with slight derivations until $e_{kin} = 10^{-6}$.
After that, no further decay can be seen. 
Numerical errors introduce the same amount of kinetic energy as it is reduced due to the viscose effects; a steady-state configuration is reached. 
The solution from the \ac{ucsph} follows the analytical solution exactly until $e_{kin} = 10^{-9}$.
After that, a similar steady-state configuration as for the \ac{wcsph} is reached.
However, the steady  state of the \ac{ucsph} solution shows a remarkable drop in the apparent kinetic energy three orders of magnitude. 
Also, for earlier times, it is visible that the kinetic energy of the \ac{ucsph} method is closer to the analytical solution than the \ac{wcsph}. 
\begin{figure}[ht]
    \centering
     \begin{subfigure}{0.48\textwidth}
        \includegraphics[width=\linewidth]{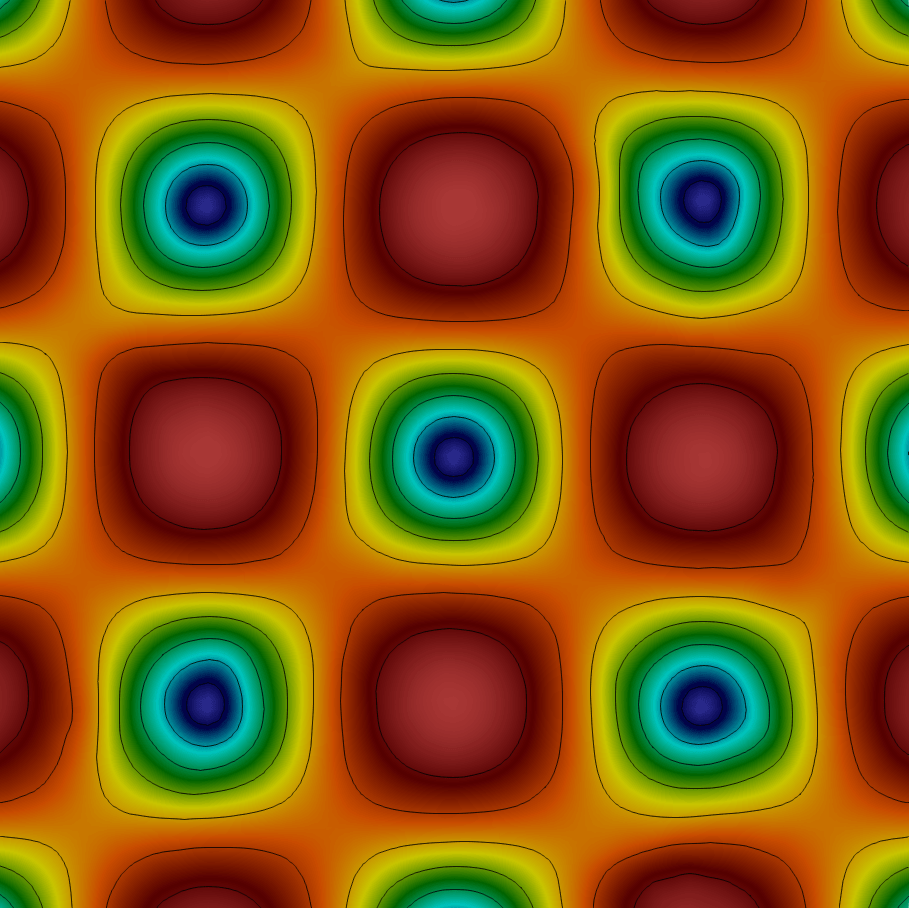}
        \caption{UCSPH}
        \label{fig:Velocity_field_at_t=5_UCSPH}
    \end{subfigure}
    \hfill
    \begin{subfigure}{0.48\textwidth}
        \includegraphics[width=\linewidth]{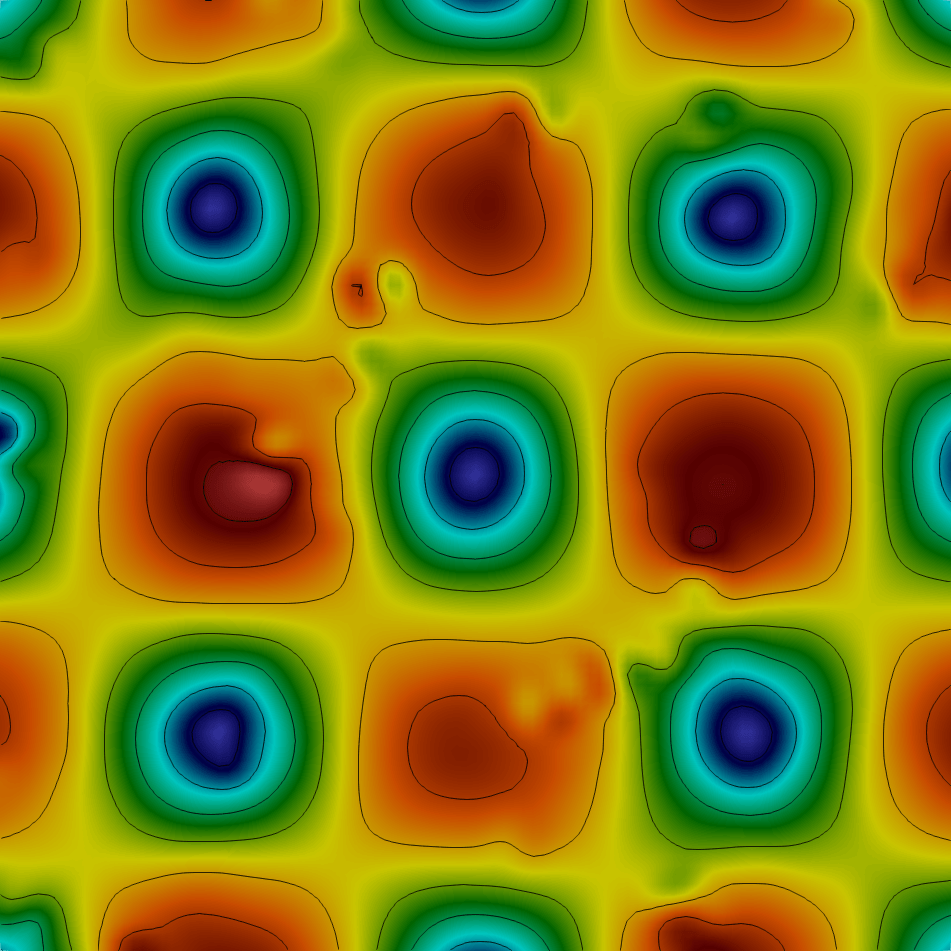}
        \caption{WCSPH}
        \label{fig:Velocity_field_at_t=5_WCSPH}
    \end{subfigure}
    \caption{The velocity magnitude field of the Taylor-Green vortex at $t = 5$.
    The color scale goes from 0 (blue) to 0.02 (red).}
    \label{fig:Velocity_field_at_t=5}
\end{figure}
\begin{figure}[ht]
    \centering
     \begin{subfigure}{0.48\textwidth}
        \includegraphics[width=\linewidth]{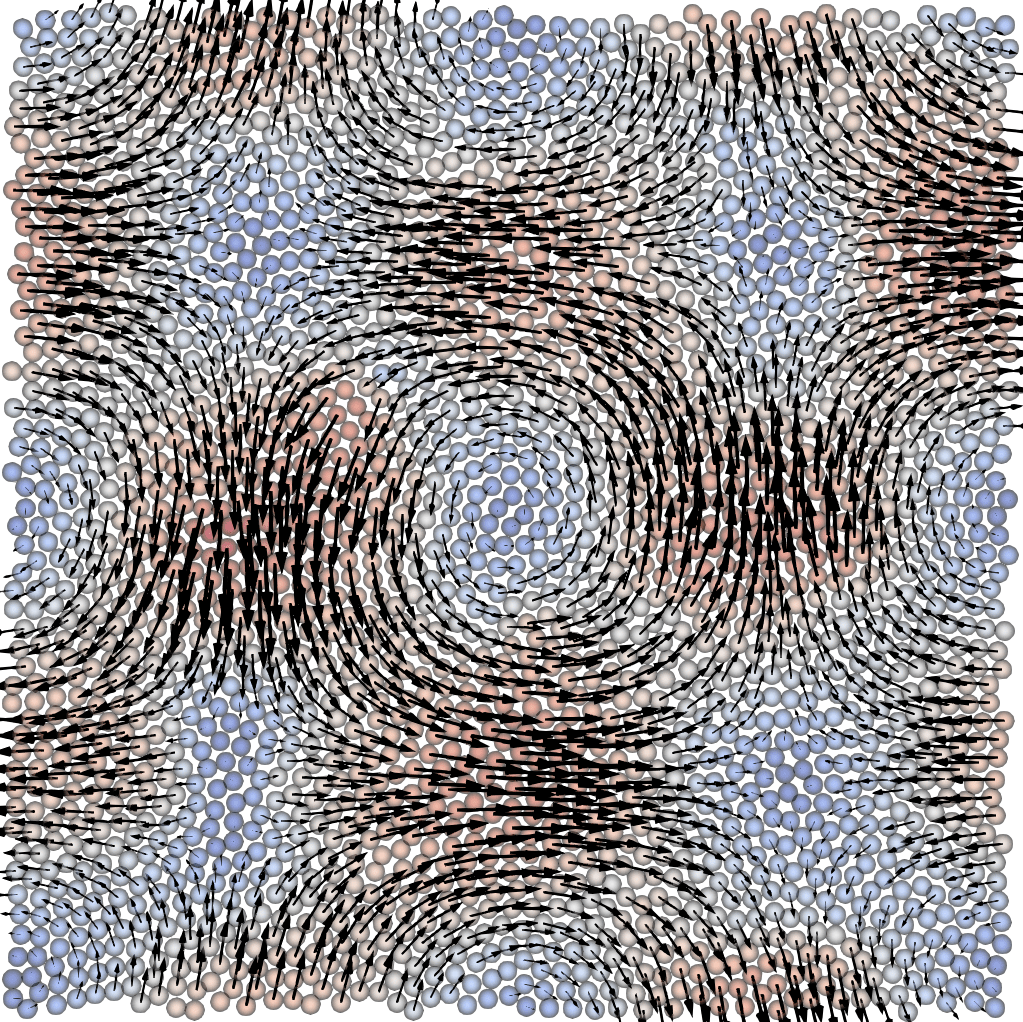}
        \caption{UCSPH}
        \label{fig:Velocity_field_at_t=10_UCSPH}
    \end{subfigure}
    \hfill
    \begin{subfigure}{0.48\textwidth}
        \includegraphics[width=\linewidth]{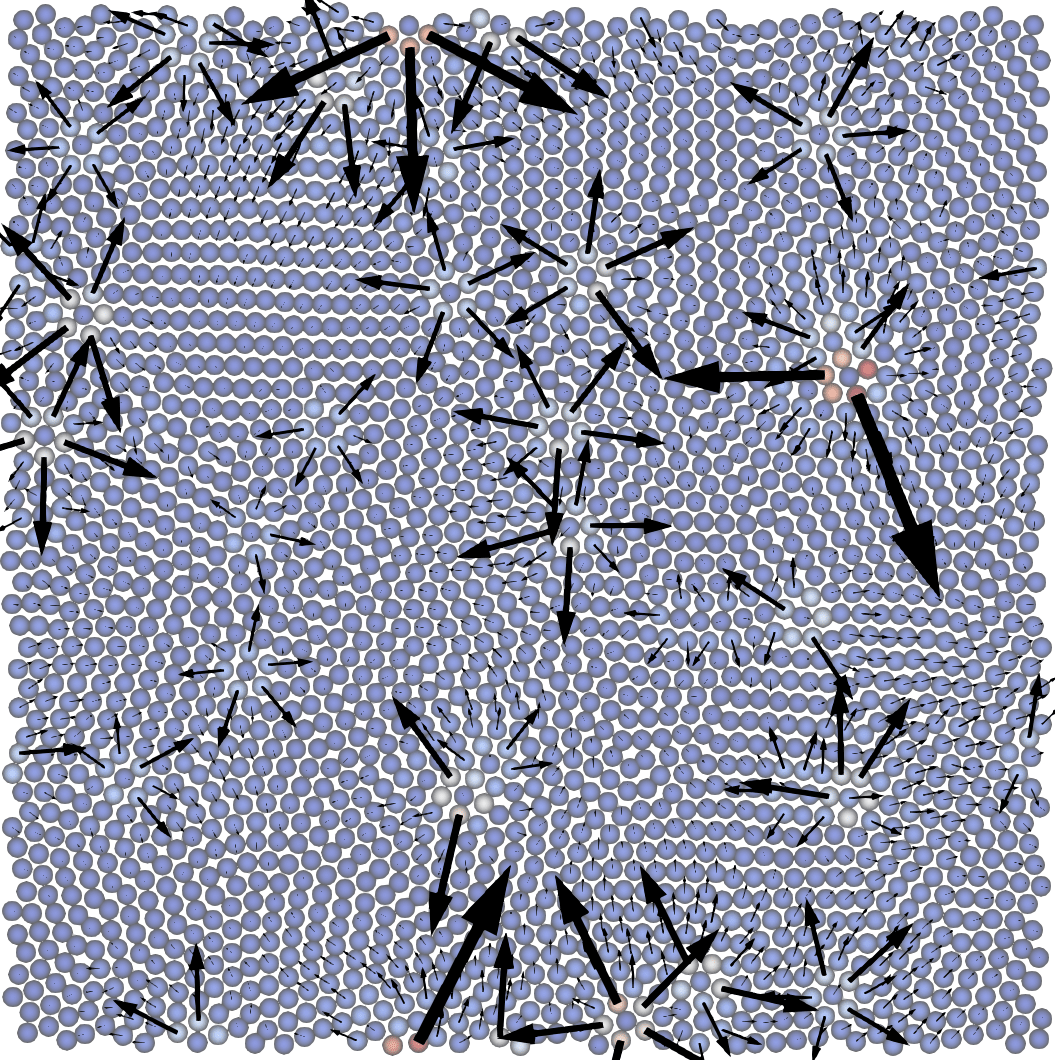}
        \caption{WCSPH}
        \label{fig:Velocity_field_at_t=10_WCSPH}
    \end{subfigure}
    \caption{The velocity vector field of the Taylor-Green vortex at $t = 10$.}
    \label{fig:Velocity_field_at_t=10}
\end{figure}
When comparing the flow fields, the same accuracy differences can be seen in Fig. \ref{fig:Velocity_field_at_t=5} and Fig. \ref{fig:Velocity_field_at_t=10}. 
In Fig. \ref{fig:Velocity_field_at_t=5}, the symmetry of the velocity magnitude fields at $t=5$ is visualized. 
For the \ac{ucsph} simulation in Fig \ref{fig:Velocity_field_at_t=5_UCSPH}, the field is almost perfectly symmetric and agrees very well with the analytical solution. 
The velocity field of the \ac{wcsph} simulation in Fig \ref{fig:Velocity_field_at_t=5_WCSPH} already shows significant asymmetries.
In Fig. \ref{fig:Velocity_field_at_t=10}, the vortex structure is visualized at $t=10$. 
The flow field of the \ac{ucsph} in Fig. \ref{fig:Velocity_field_at_t=10_UCSPH} still has a vortex structure, similar to the initial condition.
Contrary to this, the flow field of the \ac{wcsph} method in Fig. \ref{fig:Velocity_field_at_t=10_WCSPH} is dominated by noise and numerical errors - no vortex structure can be observed.

The higher accuracy of the \ac{ucsph} method has presumably two reasons. 
The first one is linked to the fact that the pressure is computed from the the density via an \ac{eos}.
Due to the kernel approximation and the particle discretization, the computed density
\begin{equation}
    \begin{aligned}
        \rho(x) 
        &= \int_{-\infty}^{\infty}\rho(\hat x) W(x-\hat x)d\hat x\\
        &=  \sum_b m_b W_{ab} + \epsilon_{\text{sph}}
    \end{aligned}
\end{equation}
contains an error $\epsilon_{\text{sph}}$ which enters the \ac{eos} via
\begin{equation}
    \begin{aligned}
        p(x) 
        &= p_0 \Big(\frac{\rho(x)}{\rho_0} - 1\Big) \\
        &= p_0 \Big(\frac{\sum_b m_b W_{ab} + \epsilon_{sph}}{\rho_0} - 1\Big)\\
        &= p_0 \Big(\frac{\sum_b m_b W_{ab}}{\rho_0} -1\Big) + p_0 \frac{\epsilon_{sph}}{\rho_0}\\
        &= p_0 \Big(\frac{\sum_b m_b W_{ab}}{\rho_0} -1\Big) + \frac{c_s^2}{\gamma}\epsilon_{sph}
        \text{.}
    \end{aligned}
\end{equation}
Clearly, the magnitude of the sound speed directly affects the error in the pressure and shall be minimized. 
The lower the pressure error, the lower the spurious accelerations, resulting in a more accurate flow field. 
A second reason for the higher accuracy of the \ac{ucsph} is the influence of $p_0$ on the transport velocity. 
For internal flows without free surfaces, the transport velocity from \citet{Adami2013}
\begin{equation}
    \tilde{v}_a(t+\Delta t) = v_a(t)+\Delta t \Big(\frac{dv_a}{dt} - \frac{p_0}{m_a} \sum_b\big(V_a^2+V_b^2\big)\frac{\partial W}{\partial r_{ab}} \Big)
    \label{eq:transport_velocity}
\end{equation}
is used.
The additional acceleration in the transport velocity formulation is scaled with $p_0$.
Similar to the previous discussion, adjusting the scaling of the transport velocity term (last term in Eq. \eqref{eq:transport_velocity}) proved beneficial and contributes to the observed enhancement of \ac{ucsph}.
The Taylor-Green vortex flow example demonstrates nicely the expected improvement in the decay of the kinetic energy. 
With adaption of $c_s$ any spurious energy injection is minimized, showing a nice vortical decay down to velocities of one order of magnitude smaller than in \ac{wcsph}.  
\begin{figure}[ht]
    \centering
    \begin{subfigure}[t]{0.495\textwidth}
        % This file was created with tikzplotlib v0.10.1.

\begin{tikzpicture}[scale=1]

\definecolor{darkgray176}{RGB}{176,176,176}
\definecolor{darkorange25512714}{RGB}{255,127,14}
\definecolor{lightgray204}{RGB}{204,204,204}
\definecolor{steelblue31119180}{RGB}{31,119,180}

\begin{axis}[
width=6.25cm,
height=6.25cm,
legend cell align={left},
legend style={fill opacity=0.8, draw opacity=1, text opacity=1, draw=lightgray204},
log basis y={10},
tick align=outside,
tick pos=left,
x grid style={darkgray176},
xlabel={Time $t$ [s]},
xmin=-0.987747864288385, xmax=20.9828435018612,
xtick style={color=black},
y grid style={darkgray176},
ylabel={Sound speed ratio},
ymode=log,
ytick style={color=black},
ytick={0.0001, 0.001,0.01,0.1,1},
yticklabels={
  \(\displaystyle {10^{4}}\),
  \(\displaystyle {10^{3}}\),
  \(\displaystyle {10^{2}}\),
  \(\displaystyle {10^{1}}\),
  \(\displaystyle {10^{0}}\)
}
]

\addplot [semithick, black] 
table [x expr=\thisrow{X}, y expr=\thisrow{P0}/500]{
X P0
0.0109150082222417 500
0.0216202800826179 500
0.0323282957342772 500
0.0430398597164565 500
0.0537577847279935 500
0.0644792452850821 500
0.0752039117041071 500
0.0859331553284531 500
0.0973153239254792 355.276613636146
0.111627345100456 214.944506761947
0.117603472238591 276.623574187309
0.13185395238105 276.623574187309
0.14611335086505 276.623574187309
0.160382481615074 276.623574187309
0.174657988178375 276.623574187309
0.188941615468915 276.623574187309
0.203230038053316 276.623574187309
0.217521664459725 276.623574187309
0.231817704028595 276.623574187309
0.246874549917016 207.722148130786
0.263279867934062 207.722148130786
0.27969623548198 207.722148130786
0.296125451072359 205.836620766091
0.313228442523866 176.854750628446
0.331154746994876 172.38946116116
0.349135353536879 172.38946116116
0.367124145743489 172.38946116116
0.385127713684946 172.38946116116
0.403184214927294 168.399624575381
0.421909875762648 154.260817101739
0.440984609745217 148.455298847885
0.460718385703543 142.494526750004
0.480512175794614 142.494526750004
0.500503188003596 135.171002702039
0.520830030160388 135.171002702039
0.541172845814994 135.171002702039
0.561609756031991 129.748391579618
0.582746498794976 121.492022849926
0.60447070530315 117.736863574962
0.626399971127784 114.090902442391
0.649056330012757 106.385112369717
0.671996999587449 104.957836704582
0.695240317150749 102.981744574045
0.718859373235153 99.0718800047956
0.742618380552554 99.0718800047956
0.766402081403645 98.9967193153025
0.790463963376809 96.6170433817852
0.814693106151823 94.8449145177438
0.839418227827122 88.1705510740423
0.864799048131108 82.6925598072081
0.891136416054846 80.1409016133109
0.91812190966796 75.0037866253689
0.945765341257624 71.8479350818764
0.973835420398786 71.0435832530638
1.00209183274906 69.6170687218419
1.03064398633259 68.7062112409176
1.05928028571001 68.7062112409176
1.08814086545817 65.6757801945161
1.11746045961318 65.6757801945161
1.14698653114867 62.5891688210609
1.17734360322772 59.7566215399312
1.2087914883807 56.412680865772
1.24046447820896 56.1200250058559
1.27263416840776 51.9757640502927
1.30599042725099 50.7083857764234
1.33943812722628 50.7083857764234
1.37325669779771 47.167523628949
1.40885465497413 44.2812243407623
1.44465402265917 44.2812243407623
1.48085109677484 42.1342478637808
1.51850091902661 39.1229726853941
1.55663446467283 39.1229726853941
1.59524563980528 36.4567028109249
1.63560294992 34.3329426492367
1.67635910197737 34.3329426492367
1.71715781601111 34.3065787589666
1.75854473938282 32.12719052756
1.80199607993812 29.2469783925504
1.84634193390049 28.7305871927173
1.89191982721482 27.1697247217214
1.93795201761013 26.7440199343129
1.98526500374449 25.3548009878669
2.03295517992212 25.3548009878669
2.08133164660162 23.3264381180729
2.13188303016236 22.4642771483024
2.18281151398493 21.7443633641224
2.23575995882031 20.1298932616975
2.28965019635234 19.3203384461912
2.34627653872354 17.3043740780194
2.40424096101406 17.24031646677
2.46271812515175 16.9753312433896
2.52146094765041 16.6580474811504
2.58240104642705 14.7071789248449
2.64584431781392 14.5175557901568
2.71030153914475 13.544039027225
2.77824056402926 11.9308116785526
2.84941274183422 11.4505197673389
2.92183861548008 10.8082842622913
2.99682462739961 10.1991158360028
3.07415378383199 9.45794800259478
3.15335001970349 9.42056141000134
3.23311498172091 9.038855347757
3.31625939185738 8.3091945842938
3.40314032002746 7.4237029182995
3.49429164538032 7.10482900611355
3.58719376192902 6.70572494181906
3.68401214704146 6.03678190356681
3.786730710488 5.36198111274544
3.89271383977978 5.20251625322243
4.0027472892987 4.68631239673641
4.11777836762036 4.43903696999882
4.23608155171277 4.11991897888112
4.35921854721562 3.80022068510694
4.48937253041389 3.35397575342294
4.62632699244531 3.17218054701583
4.76836178273446 2.89792667603885
4.91478767676009 2.76261635973029
5.06749262294305 2.47258856124725
5.22874366628416 2.2555069285058
5.3964101527434 2.08599455503289
5.5712453681364 1.92149390712255
5.75544021115596 1.69581573840469
5.95112860275469 1.55842744380983
6.15362353160111 1.44665811479289
6.36387497926474 1.36415548953536
6.58229376898605 1.25191976527868
6.80819096153822 1.16703887259972
7.04489035259447 1.0547618110538
7.2938299334082 0.955817444180812
7.55576931784171 0.858131388250151
7.83360160432257 0.758357474240504
8.13006477746556 0.665623635081819
8.44748762786236 0.575769416259347
8.78567337829839 0.520066221892006
9.14108610165025 0.468840242074795
9.51596834388936 0.423022401312967
9.91290412874777 0.376217175773588
10.3305403793134 0.339178165141522
10.7716447621828 0.305298992665688
11.2373512501394 0.272117669548663
11.7310664921991 0.240965264385631
12.2563967748058 0.214203767402973
12.814132931271 0.188922557591163
13.4074283824356 0.166728644474208
14.0387511585133 0.149017986331427
14.7037786906336 0.133280414445458
15.405404439654 0.120309435887889
16.14852505169 0.107257455319674
16.9326809494265 0.0959001619241039
17.7612542767437 0.0863208146581493
18.6370675272392 0.0768570411883868
20.0 0.068673883495903
};

% \addplot [semithick, darkorange25512714] table[x=x, y expr=\thisrow{p0}{\WCSPH}/ 500];
% \addlegendentry{WCSPH}

\end{axis}

\end{tikzpicture}
        \subcaption{Speed of sound ratio}
        \label{fig:tgv_p0_over_time}
    \end{subfigure}
    \begin{subfigure}[t]{0.495\textwidth}
        % This file was created with tikzplotlib v0.10.1.
\begin{tikzpicture}[scale=1]

\definecolor{darkgray176}{RGB}{176,176,176}
\definecolor{darkorange25512714}{RGB}{255,127,14}
\definecolor{lightgray204}{RGB}{204,204,204}
\definecolor{steelblue31119180}{RGB}{31,119,180}

\begin{axis}[
width=6.25cm,
height=6.25cm,
legend cell align={left},
legend style={fill opacity=0.8, draw opacity=1, text opacity=1, draw=lightgray204},
legend pos=north west, 
tick align=outside,
tick pos=left,
x grid style={darkgray176},
xlabel={Time $t$ [s]},
xmin=-0.987747864288385, xmax=20.9828435018612,
xtick style={color=black},
y grid style={darkgray176},
ylabel={Time step ratio},
ytick style={color=black},
scaled y ticks=false
]
\addplot [semithick, black] 
table [x expr=\thisrow{T}, y expr=\thisrow{DT}/0.000223542]{
T DT
0.0109153796275054 0.000214051660439504
0.0216211827701717 0.000214189600742832
0.0323307074780041 0.000214252354161399
0.0430443101138746 0.000214302735447802
0.0537612485581922 0.00021437261255546
0.0644825910163274 0.000214491165209213
0.07520754737147 0.000214525514370458
0.0859362830840877 0.000214612547532207
0.0973189277916103 0.000252809706815298
0.111400890731373 0.000295542017957387
0.126182158180869 0.000295709748298953
0.140972429477094 0.000295864325474572
0.155999988372887 0.000301761889635391
0.171091960262078 0.000301935220849324
0.186192113008495 0.000302070612891228
0.201300983738141 0.000302251514918213
0.216418474667872 0.000302382203492559
0.231544438530991 0.00030267057441053
0.247689953053943 0.000342944609237899
0.264837998406432 0.000342988140633649
0.281997002490114 0.000343408610073138
0.299172141092173 0.000343561860210922
0.316358080098778 0.000343880686477239
0.333559080898729 0.000344106259421901
0.350768704898751 0.00034432671863543
0.36817482475266 0.000353131895019754
0.38614647051434 0.000361294770818321
0.404393945742412 0.000372637487501448
0.423061535627419 0.000377338900534959
0.442397655338167 0.000388835722618856
0.462097213396255 0.000399687894021522
0.48211416408613 0.000400460325062557
0.502146984688936 0.000400857956504863
0.522194032662713 0.000401078296165417
0.542287843099496 0.000402422120674797
0.563024298341462 0.000423398812104158
0.58420501997727 0.000423873649198834
0.605467526580929 0.000432295336914703
0.627333139808528 0.000442620014204805
0.649483042670211 0.000443440778944752
0.67166586235714 0.000443992825143614
0.693873810290109 0.000444394659173544
0.716121683478798 0.000445776731471093
0.738838845011774 0.000455903778197974
0.761668564446307 0.000460474471404689
0.785175996304182 0.000472759893628689
0.808833928609522 0.000473427631664884
0.832522393794536 0.000474692112919957
0.856262581898588 0.00047493780073598
0.880375581820006 0.000490529539667384
0.904915589491229 0.000491101620889931
0.929963410656732 0.000513949745499788
0.95612574150382 0.000527070398739753
0.982518596029584 0.00053059763732868
1.00930005588715 0.000536469960950157
1.03618803220809 0.000543429722946342
1.06368029367097 0.000551520511501306
1.09127874446714 0.000552258563858331
1.11935572078402 0.000577973205075531
1.14911271965766 0.000603296623070169
1.17929480343533 0.000604064118224237
1.20953035481948 0.000608792071655943
1.24053257201492 0.000623173704329435
1.27170937482394 0.000623902541691403
1.30344218684052 0.000651628177162698
1.3364019498154 0.00066067330483289
1.36968355343341 0.000673429136855099
1.40370727327932 0.000695530004363371
1.43870294859453 0.00070636877792555
1.47420934238027 0.000712648473495937
1.51007381960162 0.000720288856217185
1.54613765607916 0.000725648753047253
1.58331191111014 0.000750187569887363
1.62126294811323 0.000772944313048279
1.66031282983852 0.00079330693485239
1.70001207717269 0.000794545274913802
1.7397724500603 0.000795709096538598
1.78028342289518 0.00083507242439627
1.82223326514482 0.000839650711262134
1.86490800168122 0.00087075230999257
1.90939591509671 0.000898536547359682
1.95436968293715 0.000900326007406189
1.99941758197107 0.000901467936552469
2.04504390641572 0.000934973397008729
2.09325207575612 0.00097571401609295
2.14207520549999 0.000977231776216125
2.19129889181333 0.0010036184578843
2.2428419868881 0.00104035811943413
2.29564564672733 0.00108194582520923
2.35159523868915 0.00114536751663098
2.40910108379977 0.00115457957117434
2.46687741405364 0.0011565405187677
2.52536786626557 0.00119800969046191
2.58643372446988 0.00122867338542295
2.64800887233642 0.00124320339509602
2.71190476430048 0.00129820292260252
2.77848908883084 0.00137093260386504
2.84846371995309 0.00140730261272115
2.91981235900006 0.00143326958568995
2.99153887601486 0.00143581137683308
3.06429333853898 0.00147066189981546
3.13866466405657 0.00152966120023588
3.21711260217064 0.00158760185247038
3.29656592516925 0.00159434450590909
3.3783931955442 0.00168413412852218
3.46356971151307 0.00173836664903096
3.55166435327118 0.00179253418738993
3.64174831759772 0.0018063015182506
3.73262037142684 0.00183248471202321
3.82625355979181 0.00192657717667567
3.92535283754832 0.00204292442967178
4.03090172576331 0.00216898657789774
4.14153646421769 0.00222611826816467
4.25372764707896 0.00229166680614248
4.37120103267103 0.00239247926229726
4.49381060260118 0.00252689744472542
4.62320091447354 0.00263967449186453
4.75551630820246 0.00265020558638457
4.88947517259273 0.00271741686241332
5.02915999493642 0.00285942468511182
5.1763418531281 0.00302409244475753
5.32949982563727 0.00306970964187747
5.48347703872925 0.00311646662979724
5.64418080328403 0.00330114809977564
5.81220377662493 0.00342113048153965
5.98816952016793 0.00360700354762405
6.17266272055676 0.00375911555740868
6.36312720700298 0.00388001988788645
6.5623236212839 0.00410086614393294
6.77423861813647 0.00436751409949496
6.99955331123851 0.00463627495706009
7.23810976412792 0.00488345627084852
7.48723124251358 0.00510170741821999
7.74962972577559 0.00538885004279384
8.02658910459626 0.00568366490656852
8.31558518569618 0.00590452401572454
8.61849422477505 0.00620651376541936
8.93615614499547 0.00649336431699687
9.26976927752698 0.0068525842738755
9.62244298693454 0.00724142697211493
9.99585699382181 0.00768801898162114
10.3932143964042 0.00817827507295139
10.815504441473 0.00865222938470509
11.2602151040443 0.0091475456488656
11.7311939822368 0.00968691810044176
12.2305357403813 0.0101969086176929
12.7518861188978 0.0106793870512971
13.3003988583013 0.0112520896623418
13.8779164825126 0.0118722735363149
14.486645694726 0.0124924004146306
15.1285440671818 0.0131761601866024
15.8079324133598 0.0140060129510443
16.5286761162438 0.0148547022769558
17.2906128129959 0.0156395047433907
18.0950380820908 0.0165411271459985
18.9471304383842 0.0175225724692793
};
\end{axis}

\end{tikzpicture}
        \subcaption{Time step ratio}
        \label{fig:tgv_time_step_over_time}
    \end{subfigure}
    \caption{Temporal evolution of the speed of sound ratio $c_s^{\text{UCSPH}}/c_s^{\text{WCSPH}}$ and the time step ratio $\Delta t^{\text{UCSPH}}/\Delta t^{\text{WCSPH}}$ of the Taylor-Green vortex.}
\end{figure}
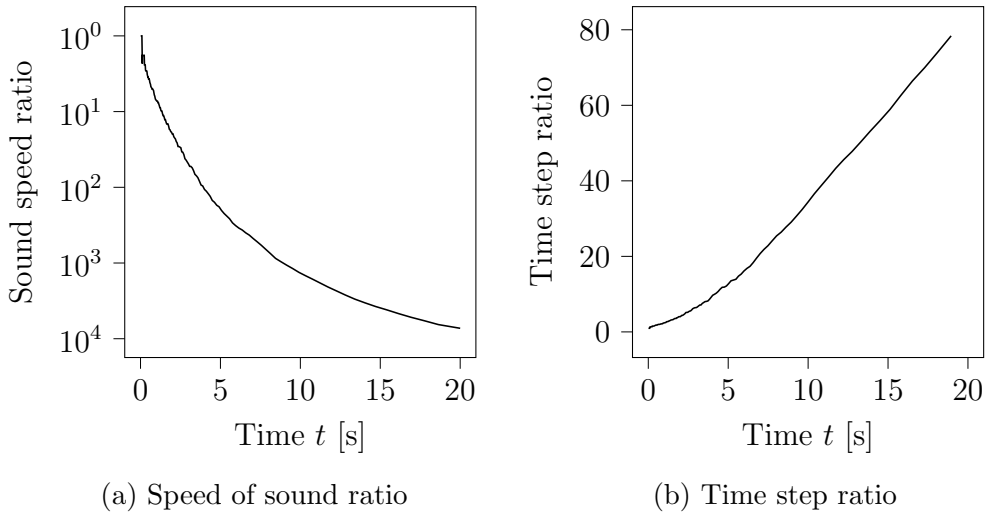

Besides improving accuracy, the \ac{ucsph} method was mainly developed to improve performance.
The Taylor-Green vortex is a perfect example to demonstrate the possible speed-up of the \ac{ucsph} method compared to the \ac{wcsph}.
In the Taylor-Green vortex case, the CFL-time step criterion is dominant.
When using \ac{wcsph}, the constant speed of sound dictates the time step in the simulation.
Since the \ac{ucsph} method adjusts the sound speed according to the decaying flow field, the CFL-time step criterion is relaxed over time. 
The sound speed ratio $c_s^{\text{UCSPH}} / c_s^{\text{WCSPH}}$ between the two methods is plotted in Fig. \ref{fig:tgv_p0_over_time}.
It can be seen that the exponential velocity decay enables an exponential decay of the sound speed in the \ac{ucsph} method as well. 
This results in an increasing time step ratio $\Delta t^{\text{UCSPH}} / \Delta t^{\text{WCSPH}}$, see Fig. \ref{fig:tgv_time_step_over_time}.
At the end of the simulation, the time step of the \ac{ucsph} method is 78 times larger than that of the \ac{wcsph} method. 
For the given case setup we observe an overall speed-up factor of $5.5$ for the \ac{ucsph} method, compared to the run time of the \ac{wcsph}.
%
%% Droplet impact 
\subsection{Falling Droplet Impact}
\label{sec:falling_droplet}
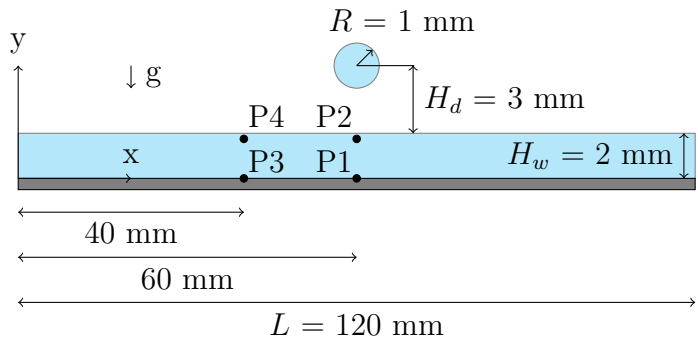
\begin{figure}[htbp]
    \centering
    \begin{tikzpicture}[scale=1.5]
        % First rectangle
        \draw[fill=gray] (-3, -0.1) rectangle (3, 0.0);
        % Second rectangle
        \draw[fill=cyan!50!white, opacity=0.5] (-3, 0.0) rectangle (3, 0.4);
        % Circle
        \draw[fill=cyan!50!white, opacity=0.5] (0.0, 1) circle [radius=0.2];
        
        % Arrows and measurements
        \draw[<->] (-3, -0.3) -- node[below] {40 mm} (-1, -0.3);
        \draw[<->] (-3, -0.7) -- node[below] {60 mm} (0.0,-0.7);
        \draw[<->] (-3, -1.1) -- node[below] {$L$ = 120 mm} (3, -1.1);
        \draw[<->] (2.9, 0) -- node[left] {$H_w$ = 2 mm} (2.9, 0.4);
        \draw (0.0, 1) -- (0.51, 1);
        \draw[<->] (0.5, 0.4) -- node[right] {$H_d$ = 3 mm} (0.5, 1);
        \draw[->] (0.0, 1) -- (0.14, 1.14);
        \node at (0.4, 1.4) {$R$ = 1 mm};
        
        \node[draw, circle, fill=black, inner sep=1pt] at (0, 0) {};
        \node at (-0.2, 0.15) {P1};
        \node[draw, circle, fill=black, inner sep=1pt] at (0, 0.35) {};
        \node at (-0.2, 0.55) {P2};
        \node[draw, circle, fill=black, inner sep=1pt] at (-1, 0) {};
        \node at (-0.8, 0.15) {P3};
        \node[draw, circle, fill=black, inner sep=1pt] at (-1, 0.35) {};
        \node at (-0.8, 0.55) {P4};
       
        % Gravity 
        \draw[->] (-2, 1) -- (-2, 0.8);
        \node at (-1.8, 0.9) {g};
        
        % coordinate system
        \draw[->] (-3, 0.0) -- (-3, 1.0);
        \node at (-3, 1.2) {y};
        \draw[->] (-3, 0.0) -- (-2, 0.0);
        \node at (-2, 0.2) {x};
        
        \end{tikzpicture}
    \caption{The initial condition for the falling droplet case.}
  \label{fig:falling_drople_initial_condirion}
\end{figure}

In the falling droplet case, a water droplet with a radius of $R=1$ mm is placed above a shallow water reservoir with a height of $H_w=2$ mm.
The distance from the water droplet to the free surface is $H_d=3$ mm.
Initially, the water droplet is at rest. 
Due to gravity, the droplet accelerates and impacts the reservoir after some time. 
The domain has a length of  $L=120$ mm in the $x$-direction and ends with periodic boundaries in that direction. 
The initial particle spacing $\Delta x$ is set to $0.1$ mm.
For post-processing, four probes are placed at $P1 = (60.0, 0.0)$, $P2 = (60.0, 1.75)$, $P3 = (40, 0.0)$, $P4 = (40, 1.75)$. 
A detailed sketch of the initial condition is given in Fig. \ref{fig:falling_drople_initial_condirion}.
Due to the impact, the falling droplet case is an interesting test case for the \ac{ucsph} method since the pressure and velocity magnitudes change significantly over time.
This offers the potential for performance increases when using a variable speed of sound formulation.
In \ac{wcsph}, the speed of sound is initialized based on the maximal pressure $p_{max}=150$ Pa and the maximal velocity $v_{max} = 0.3$ m/s, such that $c_s =  3.873$ m/s. 
The values are chosen to fulfill the weakly compressible assumptions at each time step. 
The first advantage of the \ac{ucsph} method becomes clear during the initialization of the simulation.
In \ac{wcsph}, the maximal velocity and pressure must be known for the initialization. 
Determining these values can be challenging for impact simulation and other violent flow scenarios.
If the user does not know the correct maximum values, it is even possible that the simulation will have to be repeated.
This problem does not occur with \ac{ucsph} because only the initial pressure and velocity values are required a priori.
For the \ac{ucsph} method, the initial speed of sound of the falling droplet case is computed with the initial velocity $v_{init} = 0.0$ m/s and the initial pressure $p_{init} =  40.0$ Pa, which results in a sound speed $c_s = 2$ m/s. 

\begin{figure}[hbt]
    \centering
    \begin{subfigure}{.495\linewidth}
        \centering
        \includegraphics[width=1\columnwidth]{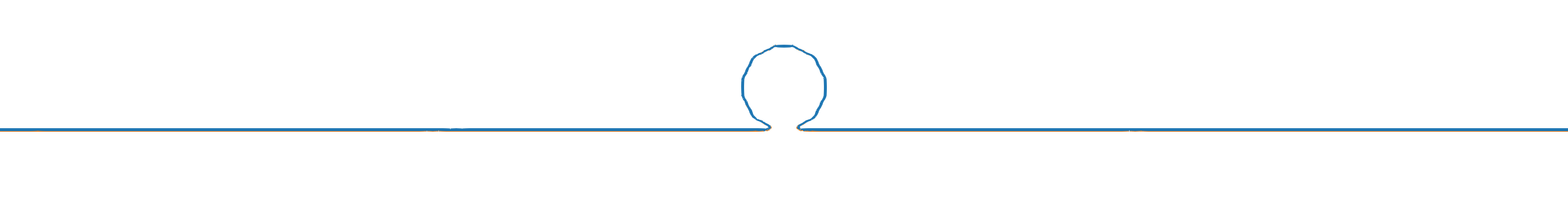}
        \caption{$t = 0.02$}\label{fig:falling_droplet_free_surface_evolution_t=0.02}
    \end{subfigure}
    \hfill
    \begin{subfigure}{.495\linewidth}
        \centering
        \includegraphics[width=1\columnwidth]{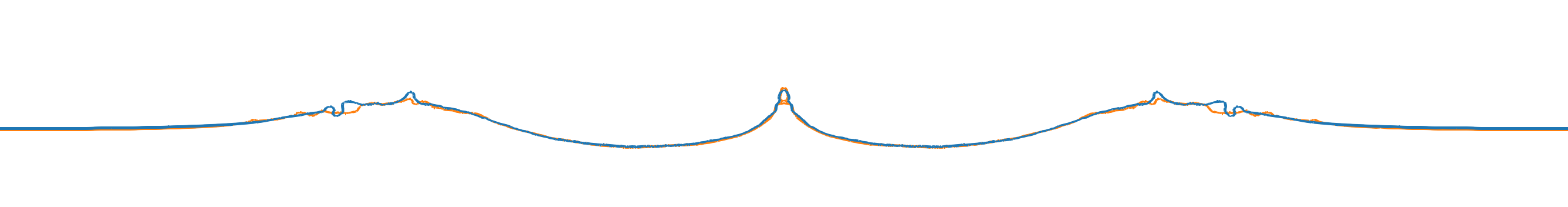}
        \caption{$t = 0.08$}\label{fig:falling_droplet_free_surface_evolution_t=0.03}
    \end{subfigure}
    \begin{subfigure}{.495\linewidth}
        \centering
        \includegraphics[width=1\columnwidth]{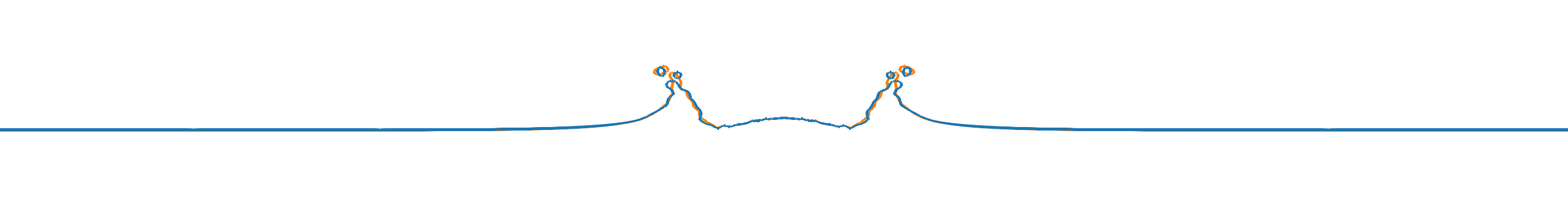}
        \caption{$t = 0.03$}\label{fig:falling_droplet_free_surface_evolution_t=0.04}
    \end{subfigure}
    \hfill
    \begin{subfigure}{.495\linewidth}
        \centering
        \includegraphics[width=1\columnwidth]{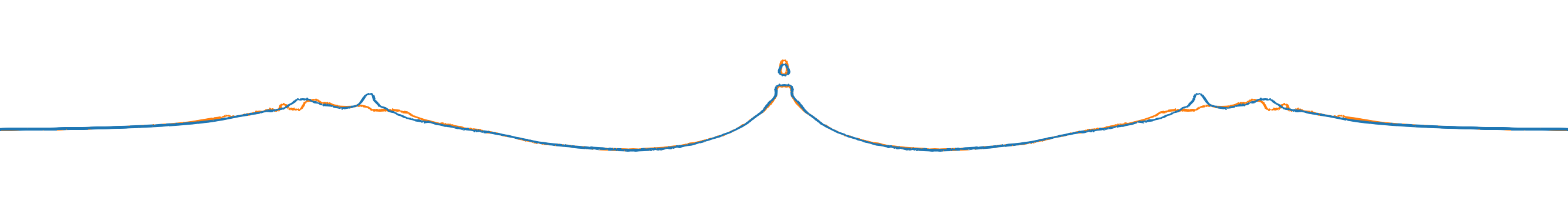}
        \caption{$t = 0.09$}\label{fig:falling_droplet_free_surface_evolution_t=0.05}
    \end{subfigure}
    \begin{subfigure}{.495\linewidth}
        \centering
        \includegraphics[width=1\columnwidth]{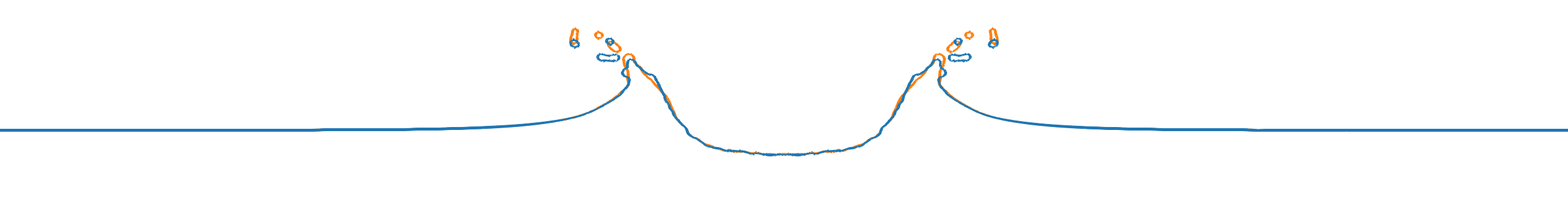}
        \caption{$t = 0.04$}\label{fig:falling_droplet_free_surface_evolution_t=0.06}
    \end{subfigure}
    \hfill
    \begin{subfigure}{.495\linewidth}
        \centering
        \includegraphics[width=1\columnwidth]{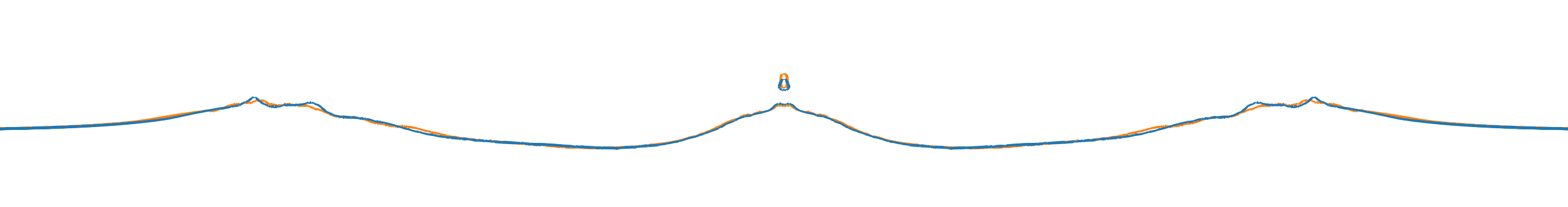}
        \caption{$t = 0.1$}\label{fig:falling_droplet_free_surface_evolution_t=0.07}
    \end{subfigure}
    \begin{subfigure}{.495\linewidth}
        \centering
        \includegraphics[width=1\columnwidth]{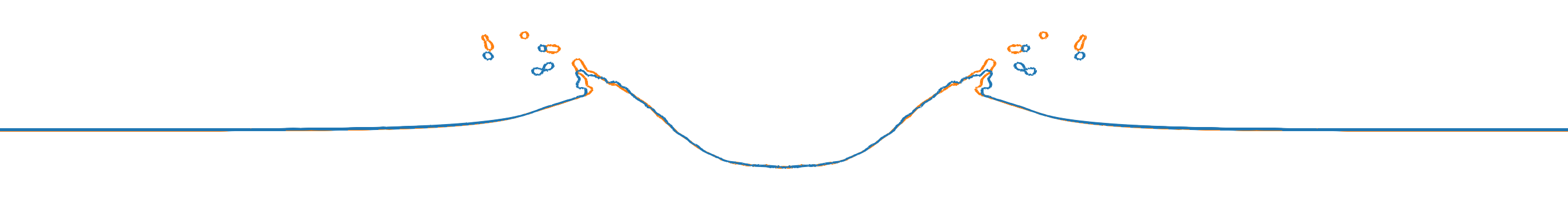}
        \caption{$t = 0.05$}\label{fig:falling_droplet_free_surface_evolution_t=0.08}
    \end{subfigure}
    \hfill
    \begin{subfigure}{.495\linewidth}
        \centering
        \includegraphics[width=1\columnwidth]{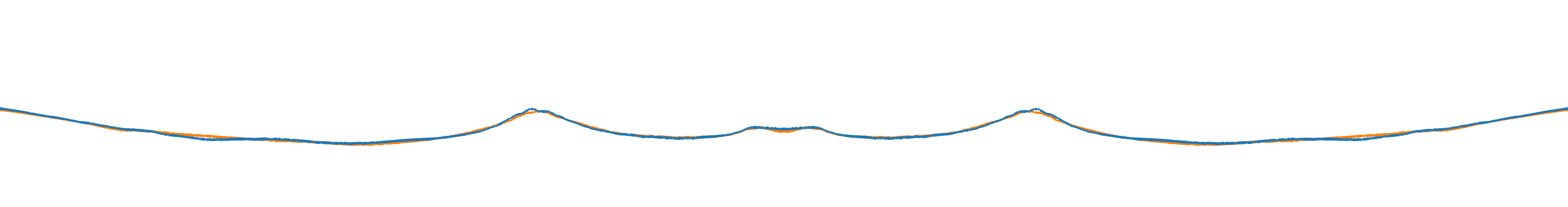}
        \caption{$t = 0.15$}\label{fig:falling_droplet_free_surface_evolution_t=0.09}
    \end{subfigure}
    \begin{subfigure}{.495\linewidth}
        \centering
        \includegraphics[width=1\columnwidth]{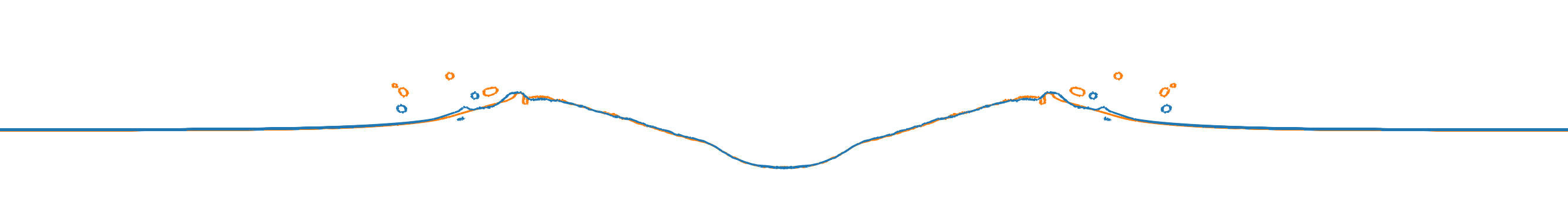}
        \caption{$t = 0.06$}\label{fig:falling_droplet_free_surface_evolution_t=0.1}
    \end{subfigure}
    \hfill
    \begin{subfigure}{.495\linewidth}
        \centering
        \includegraphics[width=1\columnwidth]{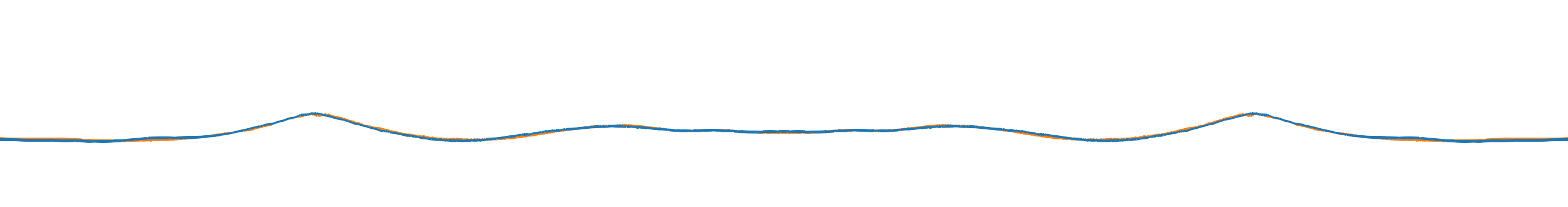}
        \caption{$t = 0.2$}\label{fig:falling_droplet_free_surface_evolution_t=0.15}
    \end{subfigure}
    \begin{subfigure}{.495\linewidth}
        \centering
        \includegraphics[width=1\columnwidth]{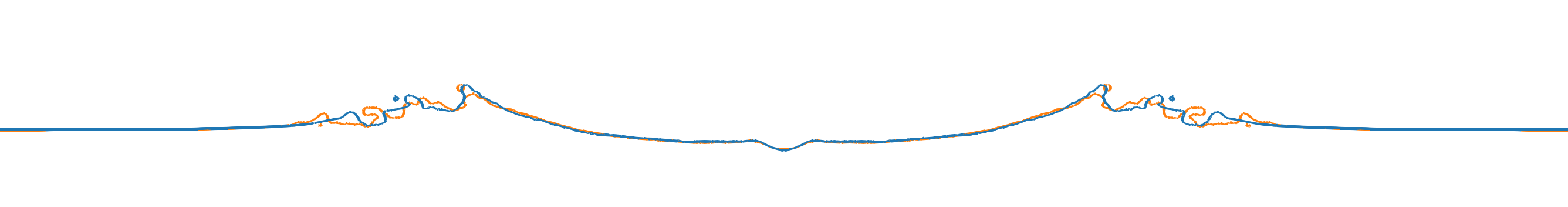}
        \caption{$t = 0.07$}\label{fig:falling_droplet_free_surface_evolution_t=0.2}
    \end{subfigure}
    \hfill
    \begin{subfigure}{.495\linewidth}
        \centering
        \includegraphics[width=1\columnwidth]{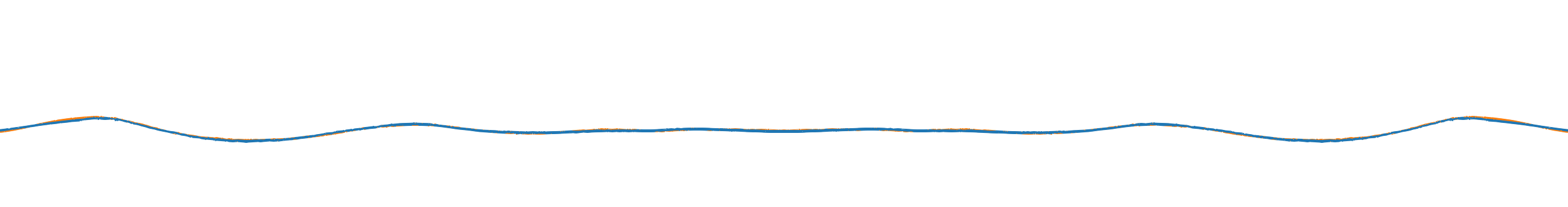}
        \caption{$t = 0.25$}\label{fig:falling_droplet_free_surface_evolution_t=0.25}
    \end{subfigure}
    \caption{The free surface evolution of the falling droplet impact. The orange line is from the WCSPH method, and the blue one is from the UCSPH method.}
    \label{fig:falling_droplet_free_surface_evolution}

\end{figure}

The temporal evolution of the free surface is shown in Fig. \ref{fig:falling_droplet_free_surface_evolution}.
The orange line is the free surface from the \ac{wcsph} method, and the blue line is the one from the \ac{ucsph} method.
Both free surfaces show an almost identical flow evolution with only marginal variations in the poorly resolved splashes directly after the droplet impact. 

\begin{figure}[htb!]
    \centering
    \includegraphics[width=.95\linewidth]{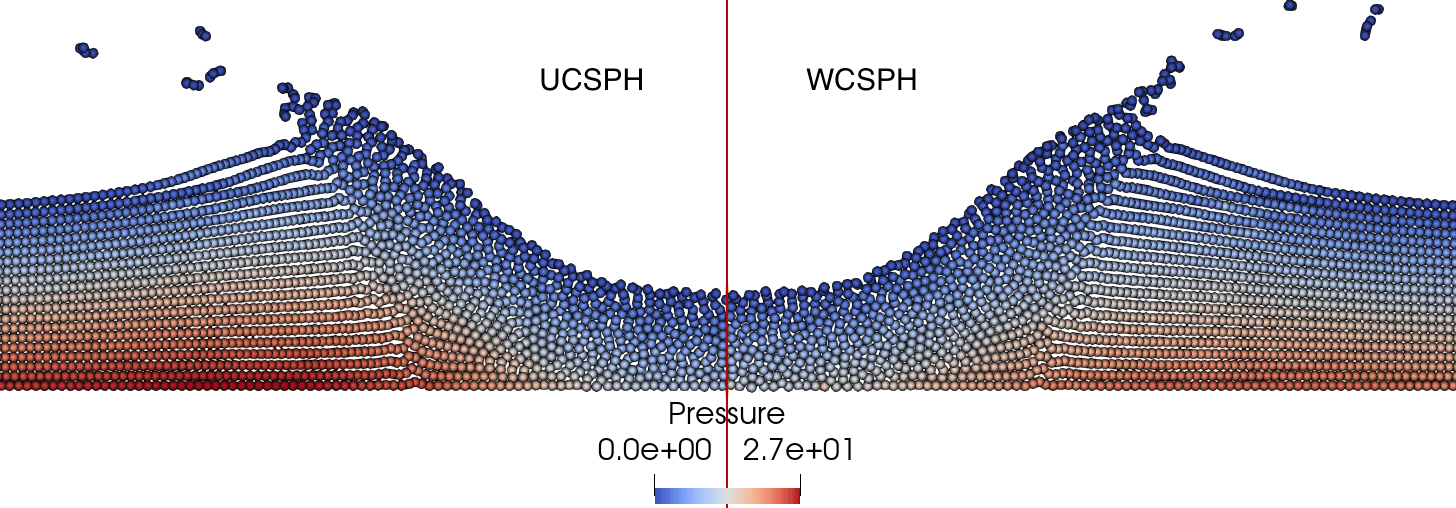}
    \caption{The pressure field from the \ac{ucsph} and the \ac{wcsph} method after the droplet impact at $t=0.05$ s.}
    \label{fig:falling_droplet_pressure_field_t005}
\end{figure}
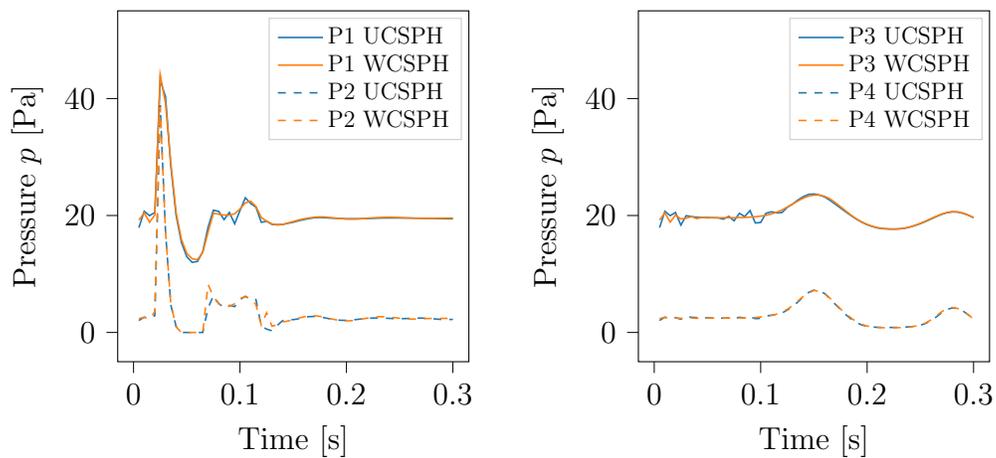
\begin{figure}[htbp]
\centering
    \begin{subfigure}[t]{.495\linewidth}
        % This file was created with tikzplotlib v0.10.1.
\begin{tikzpicture}[scale=1]

\definecolor{crimson2143940}{RGB}{214,39,40}
\definecolor{darkgray176}{RGB}{176,176,176}
\definecolor{darkorange25512714}{RGB}{255,127,14}
\definecolor{forestgreen4416044}{RGB}{44,160,44}
\definecolor{lightgray204}{RGB}{204,204,204}
\definecolor{steelblue31119180}{RGB}{31,119,180}

\begin{axis}[
width=6.25cm,
height=6.25cm,
legend cell align={left},
legend style={fill opacity=0.8, draw opacity=1, text opacity=1, draw=lightgray204, nodes={scale=0.75, transform shape}},
tick align=outside,
tick pos=left,
x grid style={darkgray176},
xlabel={Time [s]},
xmin=-0.014894721904215, xmax=0.315002636596849,
xtick style={color=black},
y grid style={darkgray176},
ylabel={Pressure $p$ [Pa]},
ymin=-5, ymax=55.0,
ytick style={color=black}
]
\addplot [semithick, steelblue31119180]
table {%
0.00500130691907567 17.9116004460502
0.010011567705972 20.7200111275484
0.0150128885650619 19.9705240950665
0.0200061018687591 20.5130269233998
0.0250038504929694 43.0817692543161
0.0300027114219987 40.3454555213822
0.0350000199512383 28.5019628363792
0.0400071448958941 19.9891464150601
0.0450013058200335 15.3892371752347
0.0500052711309716 12.9481686142803
0.055006377563618 11.9864605476382
0.0600066775285657 12.131697520912
0.0650027939834612 13.7908194289255
0.070004842184419 17.9916924980491
0.0750065547739507 20.9009951574867
0.080006546870882 20.7070487764555
0.0850030763428272 19.2899086178002
0.0900026582018142 20.5257391800557
0.0950043330738052 18.5820835683089
0.100006971118536 20.861277035232
0.105005329465969 23.0250418151452
0.110002094305075 22.0600069972184
0.115011185889413 21.4256463177458
0.120012677682081 18.8341954275836
0.125000342953679 19.0267592597555
0.130003805854794 18.4763808847777
0.135001871331348 18.4279009421651
0.1400046873274 18.4608587581446
0.145004073402984 18.6642794979786
0.15000387399518 18.8494601247181
0.155013517848305 19.0969744564739
0.160011939940416 19.3041424694504
0.165009994789123 19.4812735048841
0.170001682634074 19.6222710810579
0.175007413109685 19.6603082393576
0.180014786573849 19.6165586175269
0.185001618871682 19.5588966887289
0.190004032279123 19.5232902829562
0.195014865912494 19.4757172859391
0.200005049694421 19.4344982075732
0.205006650840583 19.4450102037138
0.210007628790606 19.433055582441
0.215009030392688 19.4524078287405
0.220014267617678 19.4841474629602
0.225006722043478 19.5168385309333
0.230001169345373 19.5483491612682
0.235012856999141 19.5701606577021
0.240010916787412 19.5804661994778
0.245010598476462 19.5795947148414
0.250011783494882 19.5753679140586
0.25501436858688 19.5703373851711
0.260002868098442 19.5639763135655
0.265007997950483 19.5490417904006
0.270014293912411 19.5285655719054
0.27500628601851 19.5097127725573
0.280014737794089 19.4958201803944
0.285008781554785 19.4726482072013
0.290003781924184 19.454431191802
0.295015120787915 19.4430576325604
0.300011926052721 19.4450842080776
};
\addlegendentry{P1 UCSPH}
\addplot [semithick, darkorange25512714]
table {%
0.00500321660553553 19.225012670157
0.0100012538755784 20.437374013092
0.0150005201195014 18.8579531861014
0.0200058207556082 20.0909413339482
0.0250036684770149 44.0701909470848
0.0300001952503032 39.2065754124409
0.0350010105483945 28.9443047153115
0.0400016093413924 20.4748301720128
0.0450011945707828 15.8845421478644
0.0500042853195055 13.529880999971
0.0550034944838409 12.6132679442795
0.0600060144717978 12.4317550336477
0.0650008700113605 13.6912035471446
0.0700046637742603 17.1467039679523
0.0750058033431135 20.3785354702734
0.0800036865008855 20.2240716904059
0.0850058462302954 20.081052529138
0.0900041246506534 20.0644599454523
0.0950053636686145 20.301804420283
0.100003462594621 21.1230534063541
0.105001095996024 22.1421495700051
0.110005958916328 22.4477324073547
0.115000987429422 21.5590124252799
0.120004998740251 19.6579543987041
0.125003951785191 19.039186572038
0.130004945611702 18.5660812496011
0.135003833707311 18.3622407525552
0.140002910380853 18.4715924682011
0.14500459109204 18.7116314623424
0.150001019431755 18.9696749492874
0.155001878009368 19.219869990399
0.160000667465374 19.4306492185323
0.165004963871839 19.5800741661265
0.170005418409832 19.7113844522336
0.175004275743213 19.7705299776926
0.180004558188784 19.7388828974471
0.185004689416183 19.6442387422293
0.190001220021152 19.5434911386861
0.19500113593041 19.449983124829
0.200002448316651 19.3932997670204
0.205003922566402 19.3722420816808
0.210000290385086 19.4007947590073
0.215000368999071 19.4591500412819
0.220003110415656 19.541401656242
0.225000318711022 19.5974739312454
0.230004721185404 19.6571935035044
0.235003495279544 19.6907059900612
0.240002992486013 19.7085804963366
0.245003174545115 19.6892325419295
0.250004006653034 19.6559155730555
0.255005457770924 19.6028264046986
0.260001084386106 19.5625285098526
0.265003637023029 19.5311548931169
0.270000299524969 19.5111014830744
0.275003846912995 19.500307791515
0.280001462716196 19.505915474142
0.285005926150125 19.5196348721455
0.290004420230218 19.5359968523207
0.295003327849627 19.5501299817654
0.300002633536194 19.5599978626953
};
\addlegendentry{P1 WCSPH}

\addplot [semithick, dashed, steelblue31119180]
table {%
0.00500130691907567 2.04970617005588
0.010011567705972 2.61397400557318
0.0150128885650619 2.45427887899973
0.0200061018687591 2.87967686508401
0.0250038504929694 38.856671800536
0.0300027114219987 17.2653853579135
0.0350000199512383 4.70318994104453
0.0400071448958941 1.01911949523562
0.0450013058200335 0
0.0500052711309716 0
0.055006377563618 0
0.0600066775285657 0
0.0650027939834612 0
0.070004842184419 4.17195700503768
0.0750065547739507 6.27695004725067
0.080006546870882 4.9260219401373
0.0850030763428272 4.30755443096844
0.0900026582018142 4.6392631184581
0.0950043330738052 4.41449940495264
0.100006971118536 5.62570313152673
0.105005329465969 6.19354318450901
0.110002094305075 5.66882030276247
0.115011185889413 5.6208796164659
0.120012677682081 1.06214815148603
0.125000342953679 0.57292955420791
0.130003805854794 0.289194571297931
0.135001871331348 1.21079827301437
0.1400046873274 2.01330261571258
0.145004073402984 2.23096050665601
0.15000387399518 2.18805746788046
0.155013517848305 2.33404789983208
0.160011939940416 2.58726082613503
0.165009994789123 2.70867007744954
0.170001682634074 2.73069238932161
0.175007413109685 2.65105982659148
0.180014786573849 2.49194805720286
0.185001618871682 2.30610658201944
0.190004032279123 2.15996450478621
0.195014865912494 2.07733649780475
0.200005049694421 2.07084323972216
0.205006650840583 2.13174734498859
0.210007628790606 2.22464399821633
0.215009030392688 2.32825557913171
0.220014267617678 2.41763077191866
0.225006722043478 2.4711157986251
0.230001169345373 2.48186285041299
0.235012856999141 2.45498337331842
0.240010916787412 2.40740918599752
0.245010598476462 2.3646023101877
0.250011783494882 2.34957485773284
0.25501436858688 2.37044484965599
0.260002868098442 2.41280458673542
0.265007997950483 2.44984096777403
0.270014293912411 2.4583924003498
0.27500628601851 2.42753327300649
0.280014737794089 2.36116610287767
0.285008781554785 2.27758086006503
0.290003781924184 2.21365499375024
0.295015120787915 2.19792867068862
0.300011926052721 2.23514243612756
};
\addlegendentry{P2 UCSPH}
\addplot [semithick, dashed, darkorange25512714]
table {%
0.00500321660553553 2.36382964951794
0.0100012538755784 2.59596841124254
0.0150005201195014 2.29553195938633
0.0200058207556082 3.56550766442529
0.0250036684770149 38.5010943129165
0.0300001952503032 17.2936871312105
0.0350010105483945 4.87043388205341
0.0400016093413924 0.925560403098567
0.0450011945707828 0
0.0500042853195055 0
0.0550034944838409 0
0.0600060144717978 0
0.0650008700113605 0
0.0700046637742603 8.22322036086224
0.0750058033431135 6.09155149728467
0.0800036865008855 4.8807590686606
0.0850058462302954 4.37558539028666
0.0900041246506534 4.46682163760926
0.0950053636686145 4.91339043577843
0.100003462594621 5.84793670754798
0.105001095996024 6.07004213919835
0.110005958916328 5.80960790382113
0.115000987429422 4.75210934695088
0.120004998740251 1.24756007795652
0.125003951785191 3.36517369309394
0.130004945611702 1.06883921559431
0.135003833707311 1.290714646772
0.140002910380853 1.78702252578547
0.14500459109204 1.95714958725495
0.150001019431755 2.17789701939904
0.155001878009368 2.43555803649065
0.160000667465374 2.6796483011234
0.165004963871839 2.83275062969987
0.170005418409832 2.85902680588822
0.175004275743213 2.75420598799374
0.180004558188784 2.56673006651217
0.185004689416183 2.35279768979641
0.190001220021152 2.16480337333448
0.19500113593041 2.03217120999066
0.200002448316651 1.97924520667789
0.205003922566402 2.01564031187333
0.210000290385086 2.13351178622294
0.215000368999071 2.29800394961668
0.220003110415656 2.47190737553911
0.225000318711022 2.60945167433177
0.230004721185404 2.68878154969639
0.235003495279544 2.69470634013415
0.240002992486013 2.63739908568966
0.245003174545115 2.53333323405923
0.250004006653034 2.41916858310607
0.255005457770924 2.32715140310137
0.260001084386106 2.28489950150901
0.265003637023029 2.29069379555478
0.270000299524969 2.32596798379322
0.275003846912995 2.36687530670873
0.280001462716196 2.39698705376667
0.285005926150125 2.40927529678837
0.290004420230218 2.407132597901
0.295003327849627 2.40108044378901
0.300002633536194 2.4021693145358
};
\addlegendentry{P2 WCSPH}

\end{axis}

\end{tikzpicture}
        % \vspace*{-8mm}
        \caption{Pressure at P1 and P2}
    \end{subfigure}
    \hfill
    \begin{subfigure}[t]{.495\linewidth}
        % This file was created with tikzplotlib v0.10.1.
\begin{tikzpicture}[scale=1]

\definecolor{crimson2143940}{RGB}{214,39,40}
\definecolor{darkgray176}{RGB}{176,176,176}
\definecolor{darkorange25512714}{RGB}{255,127,14}
\definecolor{forestgreen4416044}{RGB}{44,160,44}
\definecolor{lightgray204}{RGB}{204,204,204}
\definecolor{steelblue31119180}{RGB}{31,119,180}

\begin{axis}[
width=6.25cm,
height=6.25cm,
legend cell align={left},
legend style={fill opacity=0.8, draw opacity=1, text opacity=1, draw=lightgray204, nodes={scale=0.75, transform shape}},
tick align=outside,
tick pos=left,
x grid style={darkgray176},
xlabel={Time [s]},
xmin=-0.014894721904215, xmax=0.315002636596849,
xtick style={color=black},
y grid style={darkgray176},
ylabel={Pressure $p$ [Pa]},
ymin=-5, ymax=55,
ytick style={color=black}
]
\addplot [semithick, steelblue31119180]
table {%
0.00500130691907567 17.9116004460416
0.010011567705972 20.7200111275458
0.0150128885650619 19.9705240950767
0.0200061018687591 20.5129777475896
0.0250038504929694 18.3568767415045
0.0300027114219987 19.965259390518
0.0350000199512383 19.8068766646636
0.0400071448958941 19.6960260805694
0.0450013058200335 19.6429978538407
0.0500052711309716 19.654801468386
0.055006377563618 19.6094327698009
0.0600066775285657 19.6011221437127
0.0650027939834612 19.3777220968327
0.070004842184419 19.8523195668541
0.0750065547739507 19.0913603024523
0.080006546870882 20.3694417536092
0.0850030763428272 19.825942443692
0.0900026582018142 20.8478575141461
0.0950043330738052 18.7107849457841
0.100006971118536 18.8125519585932
0.105005329465969 20.3916518938245
0.110002094305075 20.6266377925204
0.115011185889413 20.4383601823549
0.120012677682081 20.5032194883291
0.125000342953679 21.3975577349059
0.130003805854794 21.9557128367254
0.135001871331348 22.6513899547352
0.1400046873274 23.2166064342144
0.145004073402984 23.6051121250346
0.15000387399518 23.6827521700689
0.155013517848305 23.5060098166566
0.160011939940416 23.1776848455752
0.165009994789123 22.6103208383413
0.170001682634074 21.949017740343
0.175007413109685 21.2796355787882
0.180014786573849 20.617538340686
0.185001618871682 19.9660452181313
0.190004032279123 19.3816793394826
0.195014865912494 18.8448819316008
0.200005049694421 18.4258314182806
0.205006650840583 18.1121842050039
0.210007628790606 17.8594366661248
0.215009030392688 17.7438277135377
0.220014267617678 17.6885542781068
0.225006722043478 17.6769589860296
0.230001169345373 17.7073625078074
0.235012856999141 17.8091143564823
0.240010916787412 18.0184309934594
0.245010598476462 18.2874874590633
0.250011783494882 18.6102330207912
0.25501436858688 18.9842045125434
0.260002868098442 19.3986180938789
0.265007997950483 19.8176374566749
0.270014293912411 20.1976302242903
0.27500628601851 20.4855389581386
0.280014737794089 20.6213252352769
0.285008781554785 20.583128662217
0.290003781924184 20.3808900218991
0.295015120787915 20.0440329922663
0.300011926052721 19.6275234366599
};
\addlegendentry{P3 UCSPH}
\addplot [semithick, darkorange25512714]
table {%
0.00500321660553553 19.2250126701218
0.0100012538755784 20.4373740132024
0.0150005201195014 18.8579531859605
0.0200058207556082 20.0750322846258
0.0250036684770149 19.4084120149696
0.0300001952503032 19.538801469145
0.0350010105483945 19.7977872674681
0.0400016093413924 19.4662340202513
0.0450011945707828 19.6716979569846
0.0500042853195055 19.5796217139924
0.0550034944838409 19.6293543016257
0.0600060144717978 19.5993782926713
0.0650008700113605 19.6086072494478
0.0700046637742603 19.6215516194557
0.0750058033431135 19.6199221900487
0.0800036865008855 19.660419295292
0.0850058462302954 19.6556304535679
0.0900041246506534 19.728620416533
0.0950053636686145 19.774014632634
0.100003462594621 19.8893374482675
0.105001095996024 20.0385606152145
0.110005958916328 20.2586373603528
0.115000987429422 20.5575693463982
0.120004998740251 20.9421233815082
0.125003951785191 21.4333061962807
0.130004945611702 21.9555665454262
0.135003833707311 22.4865199716337
0.140002910380853 22.9637736651861
0.14500459109204 23.3209440626964
0.150001019431755 23.4873082356779
0.155001878009368 23.4995232556595
0.160000667465374 23.2828900803103
0.165004963871839 22.8508386593903
0.170005418409832 22.2304427121363
0.175004275743213 21.5044943332226
0.180004558188784 20.7702013789809
0.185004689416183 20.0501258673555
0.190001220021152 19.407481773428
0.19500113593041 18.8891454648603
0.200002448316651 18.4861633865052
0.205003922566402 18.1484567598264
0.210000290385086 17.9182127527115
0.215000368999071 17.7841958199888
0.220003110415656 17.7060887571499
0.225000318711022 17.6824119691017
0.230004721185404 17.7209714175043
0.235003495279544 17.8224290845791
0.240002992486013 18.0412882433338
0.245003174545115 18.318685062268
0.250004006653034 18.6624573865887
0.255005457770924 19.0475337157289
0.260001084386106 19.4637775766043
0.265003637023029 19.8759077125934
0.270000299524969 20.2394743920103
0.275003846912995 20.5176834281383
0.280001462716196 20.6638472206733
0.285005926150125 20.6281936297784
0.290004420230218 20.4183028982543
0.295003327849627 20.0837574502038
0.300002633536194 19.6728934170521
};
\addlegendentry{P3 WCSPH}

\addplot [semithick, dashed, steelblue31119180]
table {%
0.00500130691907567 2.04970617005341
0.010011567705972 2.61397400557291
0.0150128885650619 2.45427887900046
0.0200061018687591 2.54717294167233
0.0250038504929694 2.26001968005598
0.0300027114219987 2.71117565875985
0.0350000199512383 2.59808706053961
0.0400071448958941 2.53275328820166
0.0450013058200335 2.49776776544399
0.0500052711309716 2.48758875819809
0.055006377563618 2.47365501116629
0.0600066775285657 2.47090591386479
0.0650027939834612 2.43180619153166
0.070004842184419 2.50415464603154
0.0750065547739507 2.38210732720917
0.080006546870882 2.59736123072142
0.0850030763428272 2.48978333667687
0.0900026582018142 2.66447654768434
0.0950043330738052 2.36117186474378
0.100006971118536 2.41379689503099
0.105005329465969 2.79562608699559
0.110002094305075 2.94622397298697
0.115011185889413 3.06868708622257
0.120012677682081 3.31045912014671
0.125000342953679 3.83863518928367
0.130003805854794 4.43411523260674
0.135001871331348 5.22415861484995
0.1400046873274 6.09323604792693
0.145004073402984 6.83695165661616
0.15000387399518 7.24685497336365
0.155013517848305 7.06177071864652
0.160011939940416 6.66744278569589
0.165009994789123 5.80633532567607
0.170001682634074 4.83874763259761
0.175007413109685 3.90352013348018
0.180014786573849 3.03892667248918
0.185001618871682 2.27465780904892
0.190004032279123 1.67168840809523
0.195014865912494 1.26948710795596
0.200005049694421 1.04666714582957
0.205006650840583 0.933414787212331
0.210007628790606 0.874376284747932
0.215009030392688 0.851106736345575
0.220014267617678 0.8486423159048
0.225006722043478 0.857990391900359
0.230001169345373 0.87702891242933
0.235012856999141 0.910425576664458
0.240010916787412 0.967383250523553
0.245010598476462 1.06606161958941
0.250011783494882 1.24011171186945
0.25501436858688 1.54352985181269
0.260002868098442 2.01996644747634
0.265007997950483 2.65592112432454
0.270014293912411 3.36730574679175
0.27500628601851 3.94431742061311
0.280014737794089 4.1771887855553
0.285008781554785 4.09209047223613
0.290003781924184 3.67849297364198
0.295015120787915 3.01857355306416
0.300011926052721 2.24320361160294
};
\addlegendentry{P4 UCSPH}
\addplot [semithick, dashed, darkorange25512714]
table {%
0.00500321660553553 2.36382964951062
0.0100012538755784 2.59596841126338
0.0150005201195014 2.29553195935818
0.0200058207556082 2.53062798659665
0.0250036684770149 2.40177000532738
0.0300001952503032 2.42865313041276
0.0350010105483945 2.47896322192562
0.0400016093413924 2.41624126857068
0.0450011945707828 2.45752501891385
0.0500042853195055 2.44135363204444
0.0550034944838409 2.45275511033448
0.0600060144717978 2.44901556342183
0.0650008700113605 2.4532433388224
0.0700046637742603 2.459105725614
0.0750058033431135 2.46370859680759
0.0800036865008855 2.47890295754495
0.0850058462302954 2.48981870828601
0.0900041246506534 2.52351088289322
0.0950053636686145 2.56296343946905
0.100003462594621 2.6341654321475
0.105001095996024 2.73977506284169
0.110005958916328 2.90136775682018
0.115000987429422 3.13966092387013
0.120004998740251 3.48012242746404
0.125003951785191 3.95637098908407
0.130004945611702 4.58397342354709
0.135003833707311 5.38105857783422
0.140002910380853 6.15459829273786
0.14500459109204 6.75157137617463
0.150001019431755 7.18527816316171
0.155001878009368 7.01744424500728
0.160000667465374 6.65259193531027
0.165004963871839 5.87040952796018
0.170005418409832 4.93618996509737
0.175004275743213 4.01389574769617
0.180004558188784 3.18120518714673
0.185004689416183 2.45420012014367
0.190001220021152 1.83189926974658
0.19500113593041 1.37778528484857
0.200002448316651 1.09887564136796
0.205003922566402 0.94172534429694
0.210000290385086 0.856738585007516
0.215000368999071 0.813673444856152
0.220003110415656 0.796975197351947
0.225000318711022 0.799834001705893
0.230004721185404 0.821243591006042
0.235003495279544 0.864276590914938
0.240002992486013 0.939944610520076
0.245003174545115 1.06721912223151
0.250004006653034 1.28337630321581
0.255005457770924 1.63417472160629
0.260001084386106 2.14916584662309
0.265003637023029 2.83618448558433
0.270000299524969 3.53888490693561
0.275003846912995 4.11098573632505
0.280001462716196 4.26940610250776
0.285005926150125 4.1050335547911
0.290004420230218 3.67495468807251
0.295003327849627 3.03222910682143
0.300002633536194 2.29298189769012
};
\addlegendentry{P4 WCSPH}

\end{axis}

\end{tikzpicture}
        % \vspace*{-8mm}
        \caption{Pressure at P3 and P4}
    \end{subfigure}
    \caption{The pressure signals of the four probes placed inside the reservoir of the falling droplet case.}
    \label{fig:falling_droplet_pressure_probes}
\end{figure}

The prediction of the free surface is essential for many engineering applications, but a reasonably accurate pressure prediction is also required.
When fluids are modeled as weakly-compressible, the artificial speed of sound influences pressure oscillations and the accuracy of the pressure in general.
Therefore, it is expected that the \ac{ucsph} method will not predict exactly the same pressure profiles as \ac{wcsph}.
However, the general pressure profiles must be captured correctly.
Fig. \ref{fig:falling_droplet_pressure_field_t005} shows the pressure field for both solvers at $t=0.05$ s. 
The two pressure fields look equivalent to each other.
For a temporal comparison, the signals from the pressure probes are compared in Fig.\ref{fig:falling_droplet_pressure_probes}.
% The pressure profiles of the four probes are plotted - the pressure from the \ac{wcsph} in orange and the one from the \ac{ucsph} in blue.
For P1 and P2, the droplet impact can be seen after $0.02$ s by the rising pressure signal.
For P3 and P4, the direct impact is not visible, but the resulting waves can be observed by a slowly increasing pressure signal.
The pressure of the \ac{ucsph} method has slightly higher fluctuations than that of the \ac{wcsph} method, which is due to the adjustment of the speed of sound.
Nevertheless, in general, the pressure signals from the two solvers are similar.

\begin{figure}
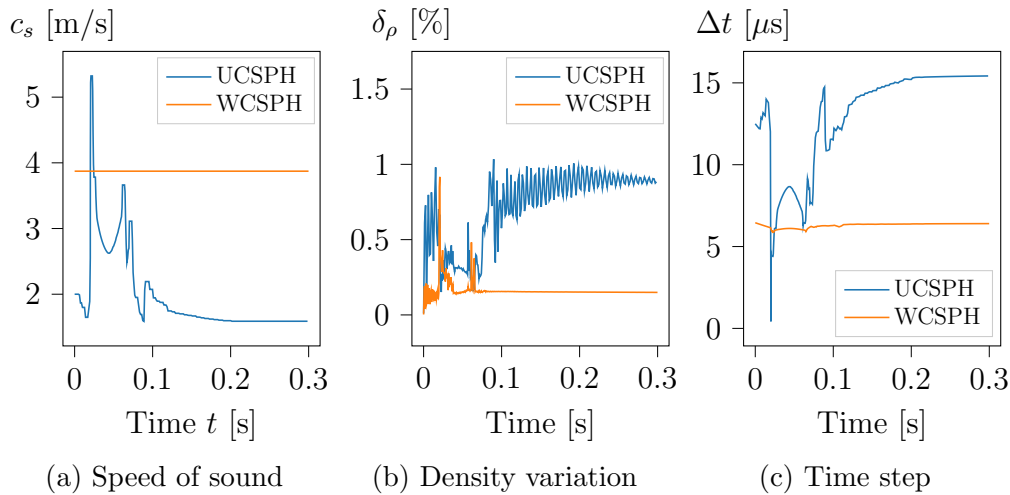

    \centering
    \begin{subfigure}[t]{.32\linewidth}
        \input{sections/results/falling_droplet/tikz_plots/p0_over_time.tex}
        \vspace*{-6mm}
        \subcaption{Speed of sound}
        \label{subfig:falling_droplet_p0}
    \end{subfigure}
    \begin{subfigure}[t]{.32\linewidth}
        \input{sections/results/falling_droplet/tikz_plots/density_variation_over_time.tex}
        \vspace*{-6mm}
        \subcaption{Density variation}
        \label{subfig:falling_droplet_density_variation}
    \end{subfigure}
    \begin{subfigure}[t]{.32\linewidth}
        \input{sections/results/falling_droplet/tikz_plots/Time_steps_over_time.tex}
        \vspace*{-6mm}
        \subcaption{Time step}
        \label{subfig:falling_droplet_time_step}
    \end{subfigure}
    
    \caption{Comparison of the temporal evolution of different solver properties from the \ac{ucsph} and the \ac{wcsph} for the falling droplet case.}
    \label{fig:falling_droplet_p0_TimeStep_DensityVariation}

\end{figure}

In Fig. \ref{subfig:falling_droplet_p0}, the time-dependent speed of sound of the \ac{ucsph} method is plotted.
The primary impact occurs at 0.02 seconds and the secondary impacts until 0.19 seconds. 
After that the maximal pressure and the maximal velocity become significantly lower. This enables a lower speed of sound while still fulfilling the weakly compressible assumptions.
In Fig. \ref{subfig:falling_droplet_density_variation}, the maximal density variation is plotted. 
For both methods, the density variation is below the admissible 1\% limit at each time step.
Nevertheless, for the \ac{wcsph} method, the maximum density variation is mostly below 0.02\%, meaning the \ac{eos} is stiffer than required.
This leads to unnecessary small time steps for that time interval. 
On the contrary, the maximal density variation from the \ac{ucsph} method is usually close to 1\%, which means the \ac{eos} is as stiff as required to fulfill the weakly compressible assumption.
This also leads to larger time steps, as shown in Fig. \ref{subfig:falling_droplet_time_step}. 
After droplet impact, the time steps of the \ac{ucsph} method are about 2.4 times larger than that of the \ac{wcsph} method. 
%
%% Dam break 
\subsection{Dam Break}
\label{sec:dam_break}
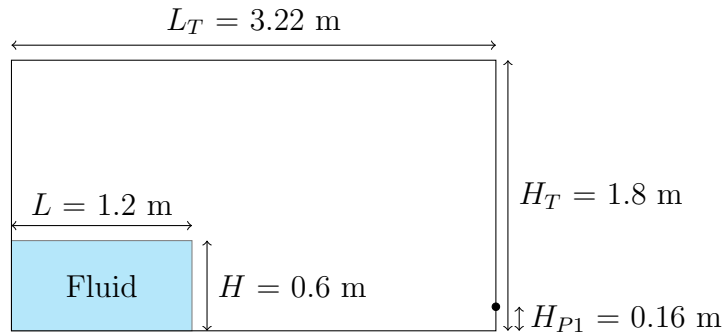
\begin{figure}[bht!]
    \centering
    \begin{tikzpicture}[scale=1.0]
        % Tank rectangle
        \draw (0.0, 0.0) rectangle (6.44, 3.6);
        % Fluid rectangle
        \draw[fill=cyan!50!white, opacity=0.5] (0.0, 0.0) rectangle (2.4, 1.2);
        \node at (1.2, 0.6) {Fluid};
        % Probe
        \node[draw, circle, fill=black, inner sep=1pt] at (6.44, 0.32) {};
        
        % % Arrows and measurements
        \draw[<->] (0.0, 3.8) -- node[above] {$L_T$ = 3.22 m} (6.44, 3.8);
        \draw[<->] (0.0, 1.4) -- node[above] {$L$ = 1.2 m} (2.4, 1.4);
        \draw[<->] (2.6, .0) -- node[right] {$H$ = 0.6 m} (2.6, 1.2);
        \draw[<->] (6.6, .0) -- node[right] {$H_T$ = 1.8 m} (6.6, 3.6);
        \draw[<->] (6.75, .0) -- node[right] {$H_{P1}$ = 0.16 m} (6.75, 0.32);
        \end{tikzpicture}
    \caption{The initial conditions of the dam break from \citet{Buchner2002}.}
  \label{fig:dam_break_initial_condirion}
\end{figure}

As last validation case, we simulated the well-known dam break case.
Our numerical set-up is equivalent to the experimental set-up from \citet{Buchner2002}, which is also frequently used for numerical experiments \cite{Colagrossi2003, Adami2012, Marrone2011}.
At the beginning of the simulation, a fluid column with height   $H=0.6$ m and length $L = 1.2$ m is separated with a gate from the rest of the domain. 
The fluid's density and viscosity are set to $\rho = 1000$ $\text{kg/m}^3$ and $\eta=0.001$ $\text{Ns/m}^2$.
The gate opens at $t=0$ s, and the fluid flows in the tank. 
The tank has a length of $L_T = 3.22$ m and a height of $H_T = 1.8$ m.
The domain is bounded with no-slip walls at both sides and the bottom. 
The walls are modeled according to \citet{Adami2012}.
\begin{figure}
    \centering
    \begin{subfigure}{.495\linewidth}
        \centering
        \includegraphics[width=1\columnwidth]{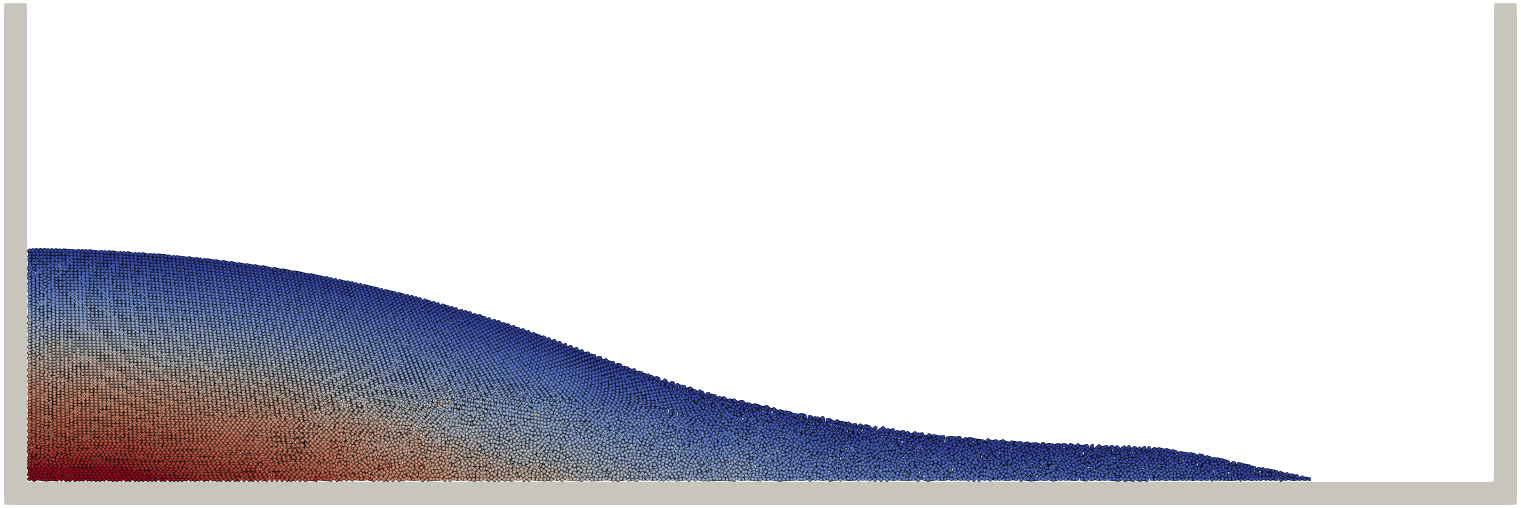}
        \caption{\ac{ucsph} at $T = 2$}\label{fig:ucsph_dam_break_evolution_t=0.5}
    \end{subfigure}
        \hfill
    \begin{subfigure}{.495\linewidth}
        \centering
        \includegraphics[width=1\columnwidth]{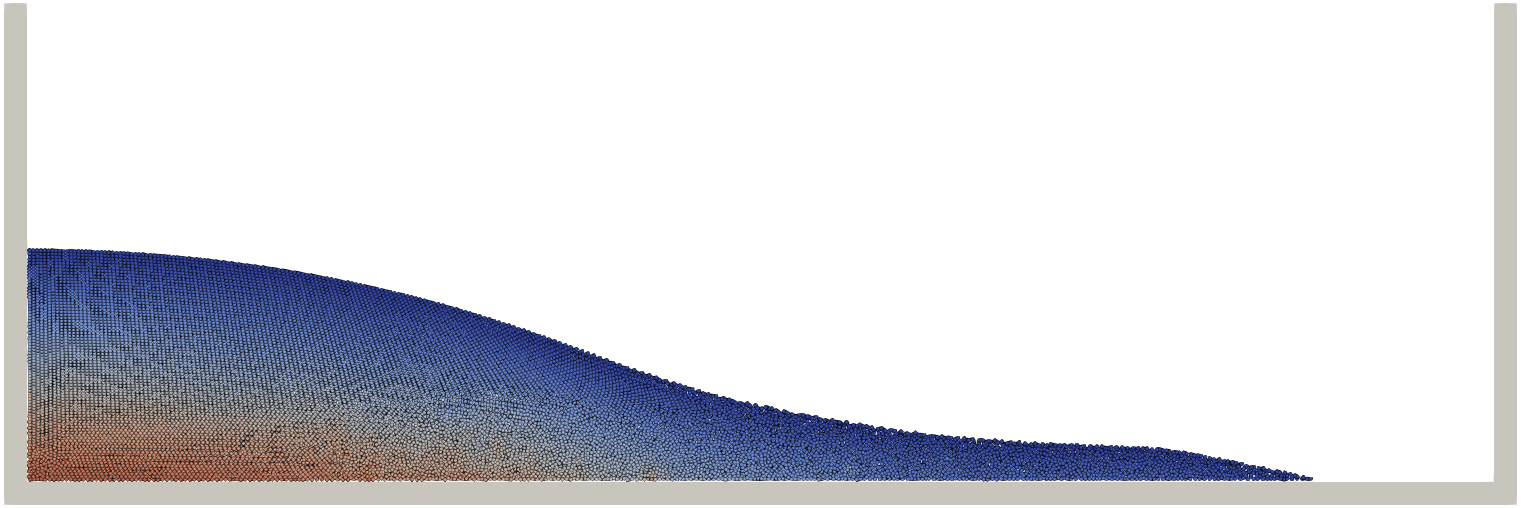}
        \caption{\ac{wcsph} at $T = 2$}\label{fig:wcsph_dam_break_evolution_t=0.5}
    \end{subfigure}
    
    \bigskip
    \begin{subfigure}{.495\linewidth}
        \centering
        \includegraphics[width=1\columnwidth]{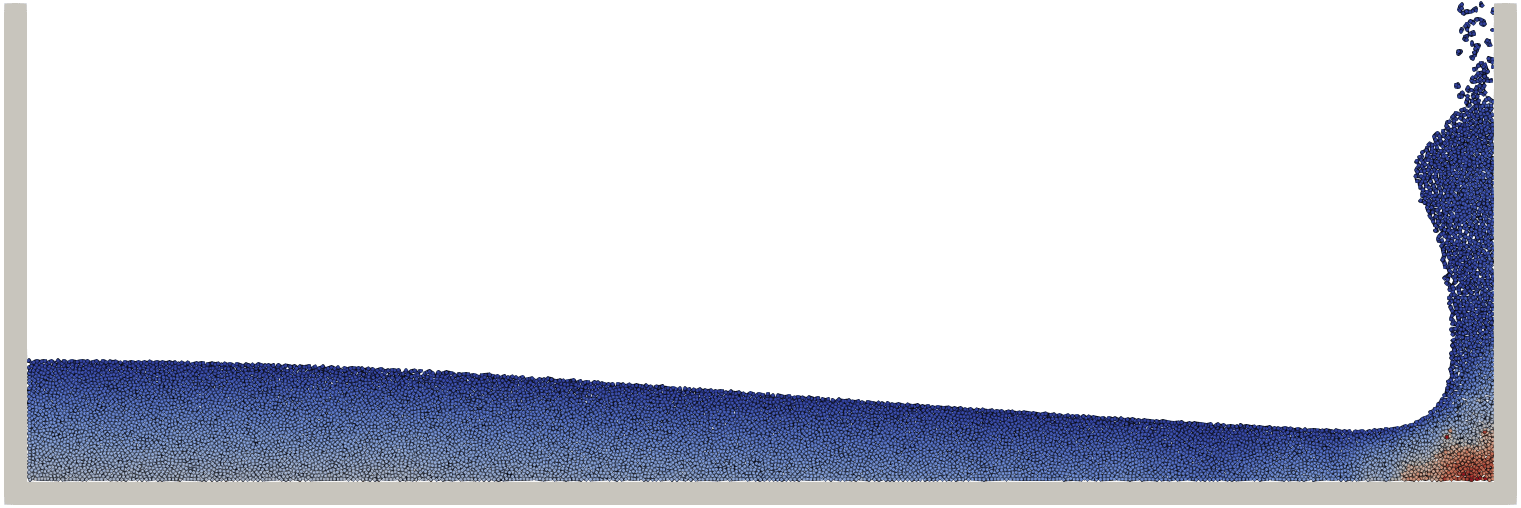}
        \caption{\ac{ucsph} at $T = 4.54$}\label{fig:ucsph_dam_break_evolution_t=1.1125}
    \end{subfigure}
    \hfill
    \begin{subfigure}{.495\linewidth}
        \centering
        \includegraphics[width=1\columnwidth]{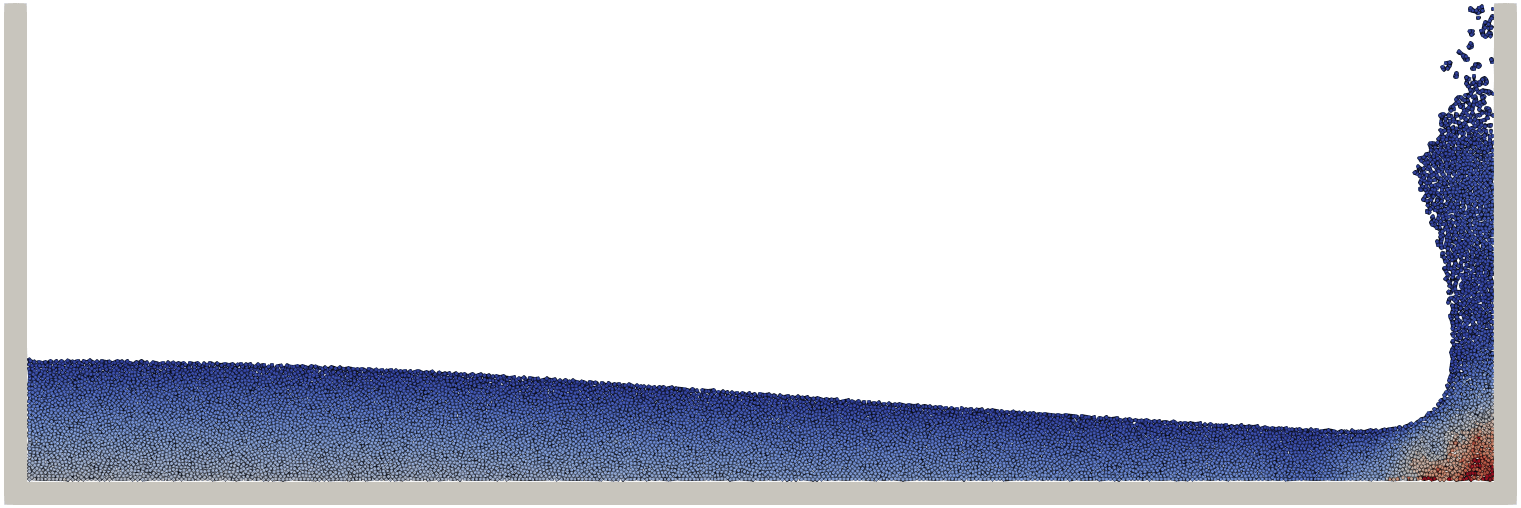}
        \caption{\ac{wcsph} at $T = 4.54$}\label{fig:wcsph_dam_break_evolution_t=1.1125}
    \end{subfigure}
    
    \bigskip
    \begin{subfigure}{.495\linewidth}
        \centering
        \includegraphics[width=1\columnwidth]{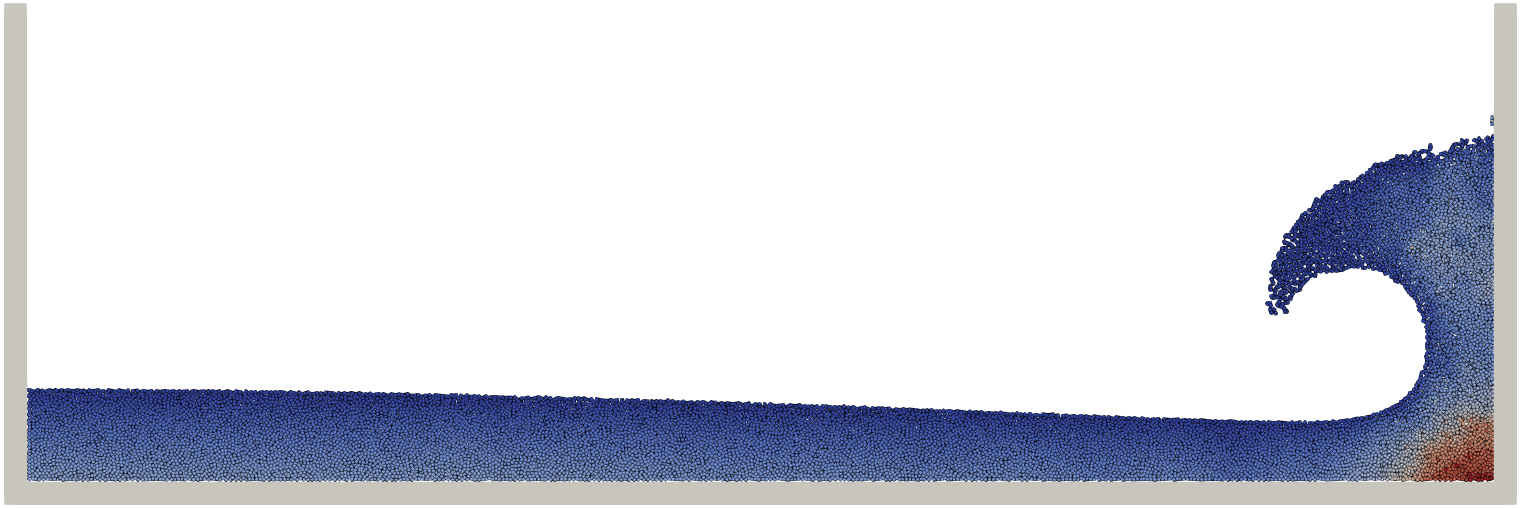}
        \caption{\ac{ucsph} at $T = 5.65$}\label{fig:ucsph_dam_break_evolution_t=1.4}
    \end{subfigure}
    \hfill
    \begin{subfigure}{.495\linewidth}
        \centering
        \includegraphics[width=1\columnwidth]{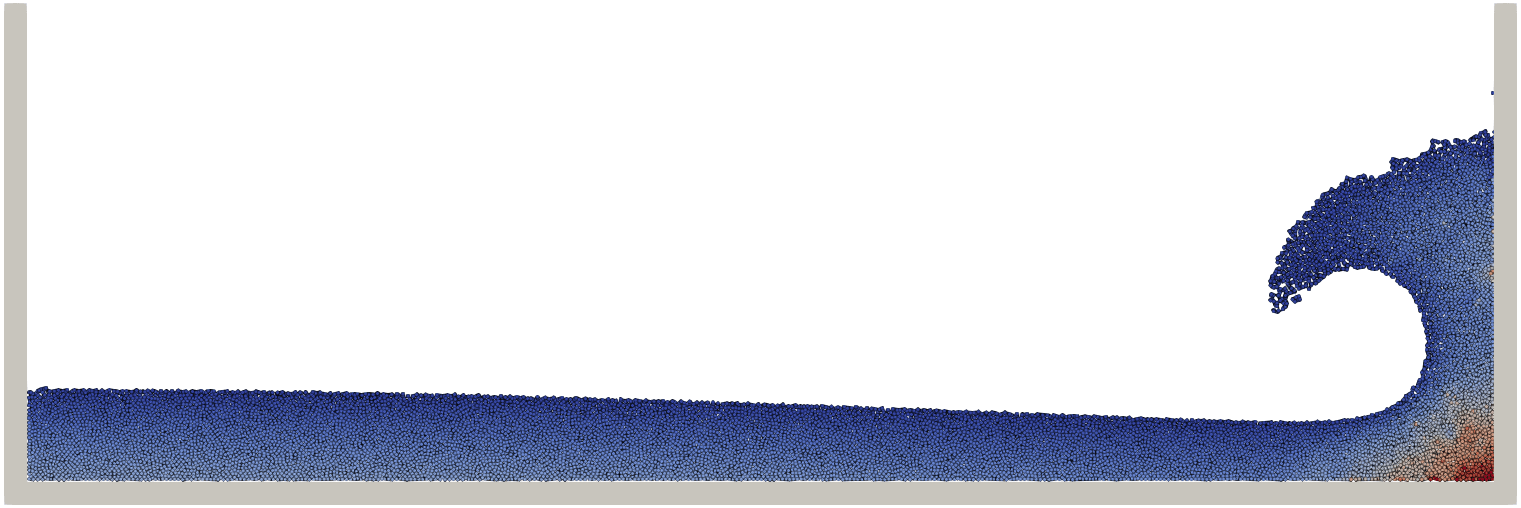}
        \caption{\ac{wcsph} at $T = 5.65$}\label{fig:wcsph_dam_break_evolution_t=1.4}
    \end{subfigure}
    
    \bigskip
    \begin{subfigure}{.495\linewidth}
        \centering
        \includegraphics[width=1\columnwidth]{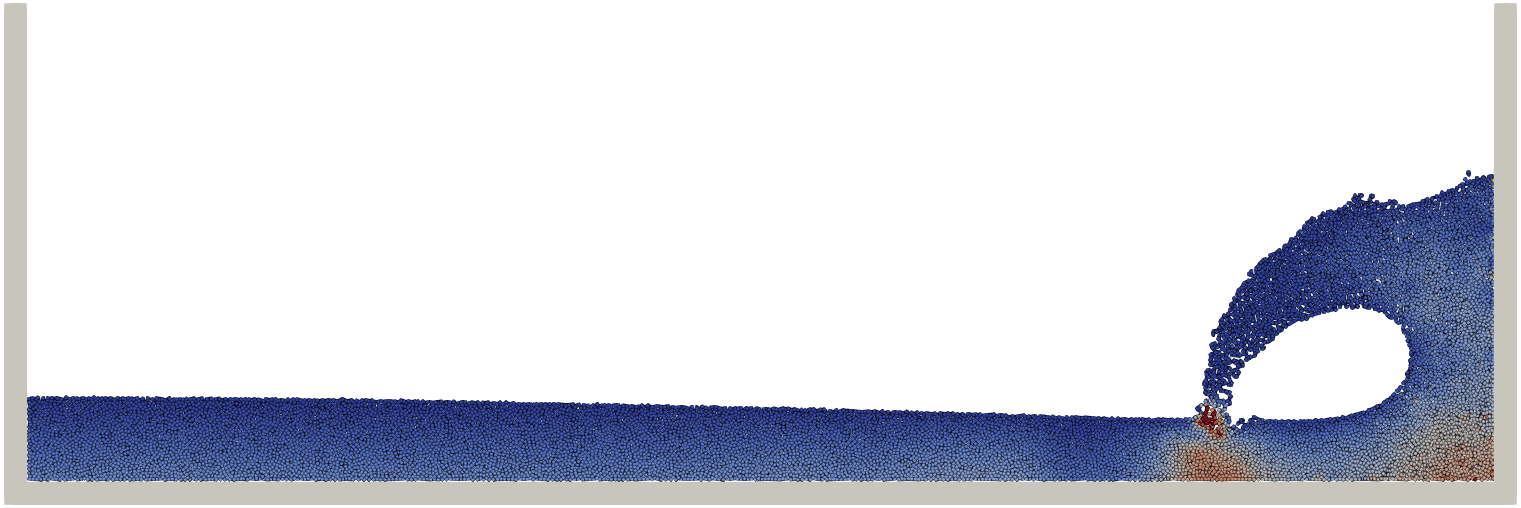}
        \caption{\ac{ucsph} at $T = 6.07$}\label{fig:ucsph_dam_break_evolution_t=1.5}
    \end{subfigure}
        \hfill
    \begin{subfigure}{.495\linewidth}
        \centering
        \includegraphics[width=1\columnwidth]{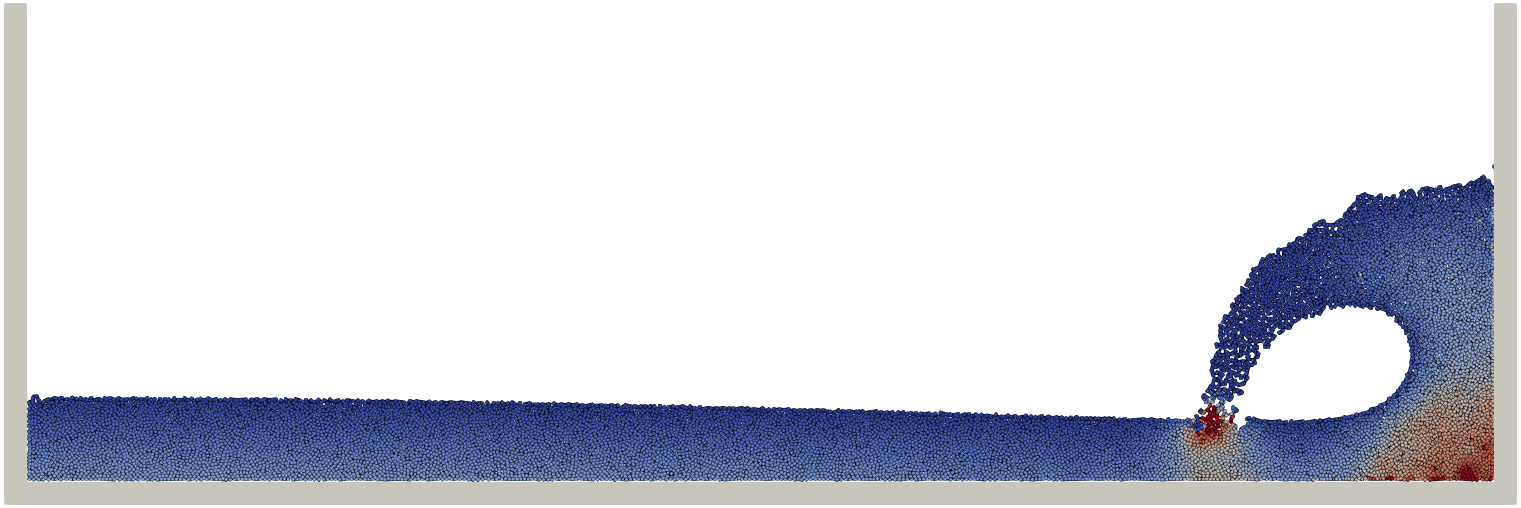}
        \caption{\ac{wcsph} at $T = 6.07$}\label{fig:wcsph_dam_break_evolution_t=1.5}
    \end{subfigure}
    
    \bigskip
    \begin{subfigure}{.495\linewidth}
        \centering
        \includegraphics[width=1\columnwidth]{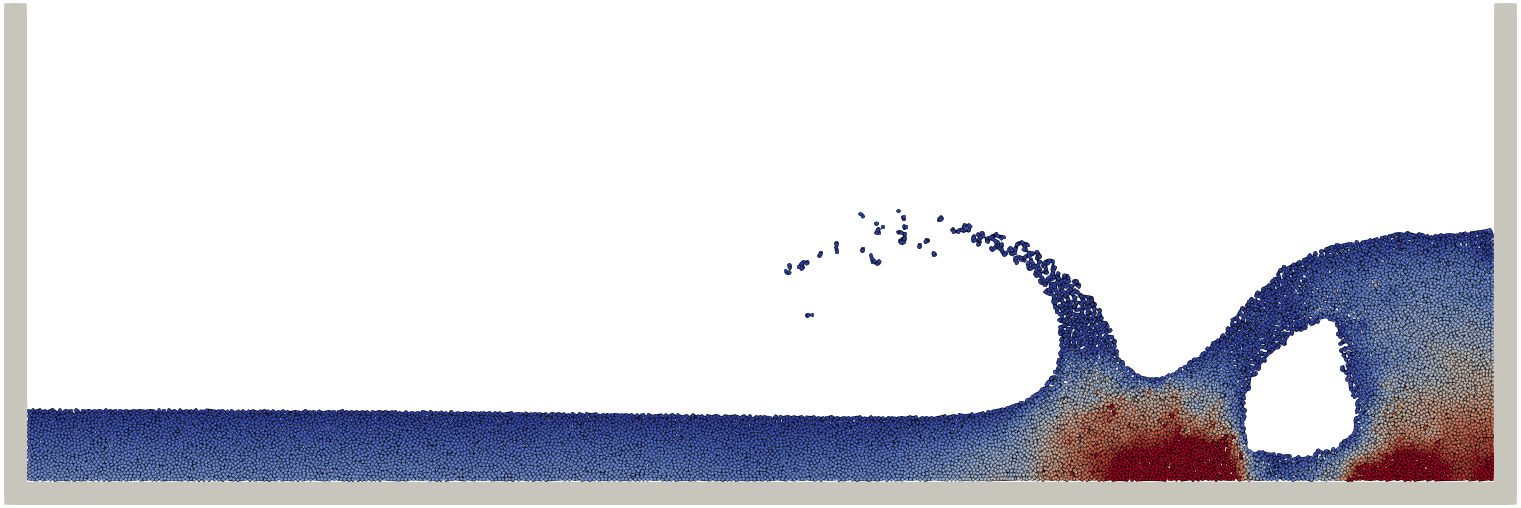}
        \caption{\ac{ucsph} at $T = 6.87$}\label{fig:ucsph_dam_break_evolution_t=1.7}
    \end{subfigure}
    \hfill
    \begin{subfigure}{.495\linewidth}
        \centering
        \includegraphics[width=1\columnwidth]{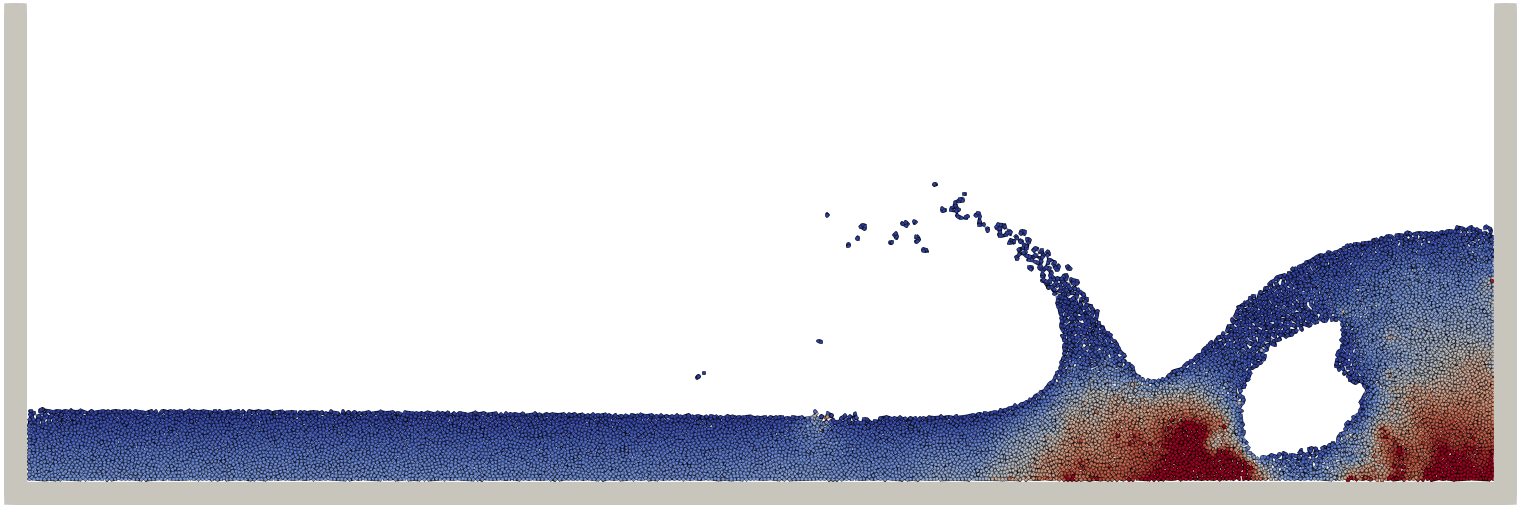}
        \caption{\ac{wcsph} at $T = 6.87$}\label{fig:wcsph_dam_break_evolution_t=1.7}
    \end{subfigure}
    
    \caption{The particle evolution of the dam break simulation, with the \ac{ucsph} (left) and the \ac{wcsph} (right). 
    The color map indicates the pressure of the fluid - blue for $p=0$ Pa and red for $p=\rho g H = 5886$ Pa.}
    \label{fig:dam_break_evolution}

\end{figure}
At the right wall, a pressure sensor is placed at a height of $H_{P1} = 0.16$ m.
The case is considered two-dimensional, and the domain is discretized with an initial particle spacing of $\Delta x = 0.00375$ m, such that $H/\Delta x = 80$.
In the \ac{ucsph} method, the initial speed of sound is based on the initial velocity $v_{init} = 0$ m/s and the initial pressure $p_{init} = 2 \rho g H = 11772$ Pa.
For the \ac{wcsph} method, the speed of sound depends on the maximal velocity $v_{max} = 6$ m/s and the maximal pressure $p_{max} = 200000$ Pa.

\begin{figure}[htbp]
    \centering
    \input{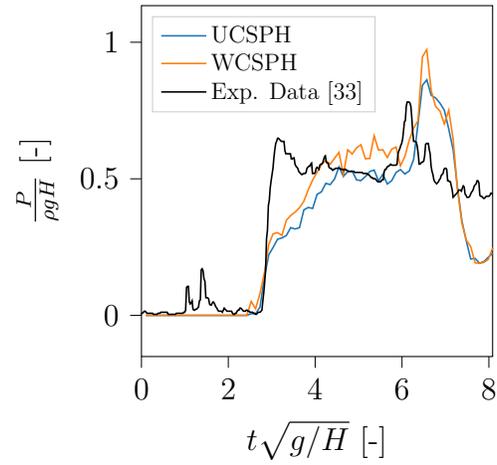}
    \caption{Pressure signal of the pressure probe P1.}
    \label{fig:dam_break_pressure_probe_P1}
\end{figure}

In Fig. \ref{fig:dam_break_evolution}, the snapshots of the two simulations are shown. 
We want to highlight the excellent agreement of both simulations. 
This demonstrates the accuracy and robustness of the \ac{ucsph} for complex cases with violent impacts.
More quantitatively, we compare the pressure signal of the probe at the right wall.
The numerical pressure signal of the probe is plotted in Fig. \ref{fig:dam_break_pressure_probe_P1}.
The figure shows that the two numerical pressure signals are close to each other.
Additionally, the experimental data from \citet{Buchner2002} are plotted.
Both numerical signals slightly differ from the experimental data.
The difference between the numerical and experimental signals is similar to the results from \citet{Adami2012}, which also used a standard \ac{wcsph} solver. 
%
%% Long dam break case
We repeat the simulation with a lower resolution ($H/\Delta x = 40$) but with a much longer simulation time of 20 seconds. 
This is done to see the solver's behavior when the fluid motion decays.
\begin{figure}
    \centering
    \begin{subfigure}[t]{1.\linewidth}
        \begin{subfigure}{.495\linewidth}
            \centering
            \includegraphics[width=1\columnwidth]{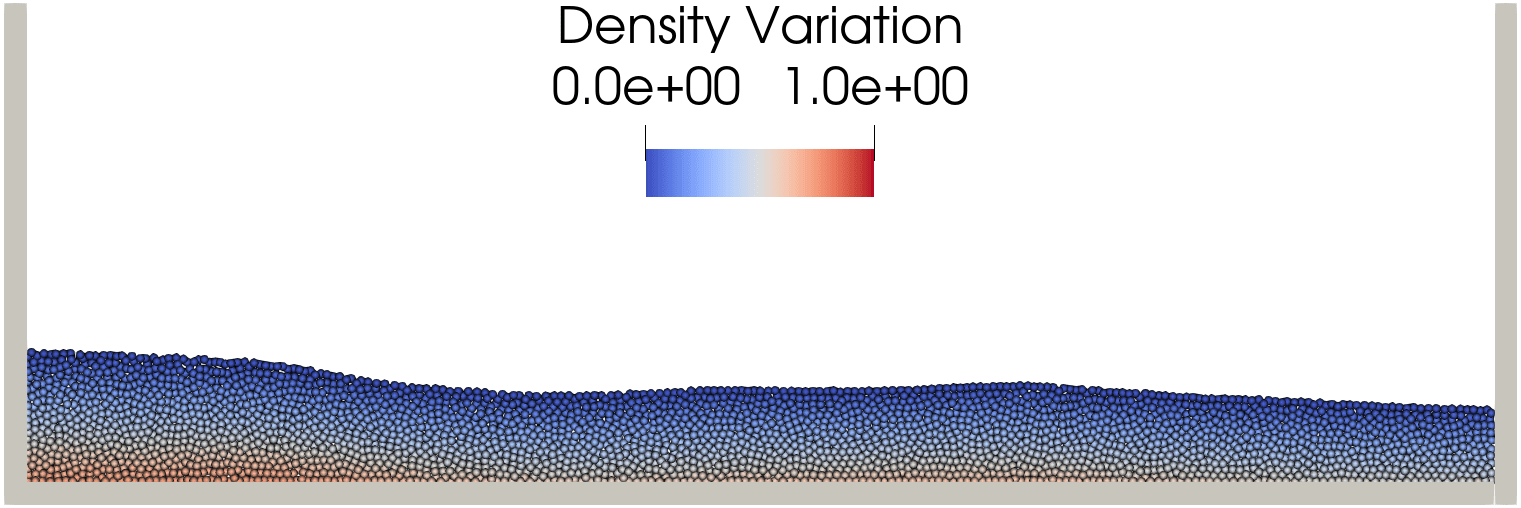}
            % \caption{Density variation of the \ac{ucsph}}
            \label{fig:UCSPH_dam_break_evolution_pressure_t=20}
        \end{subfigure}
        \begin{subfigure}{.495\linewidth}
            \centering
            \includegraphics[width=1\columnwidth]{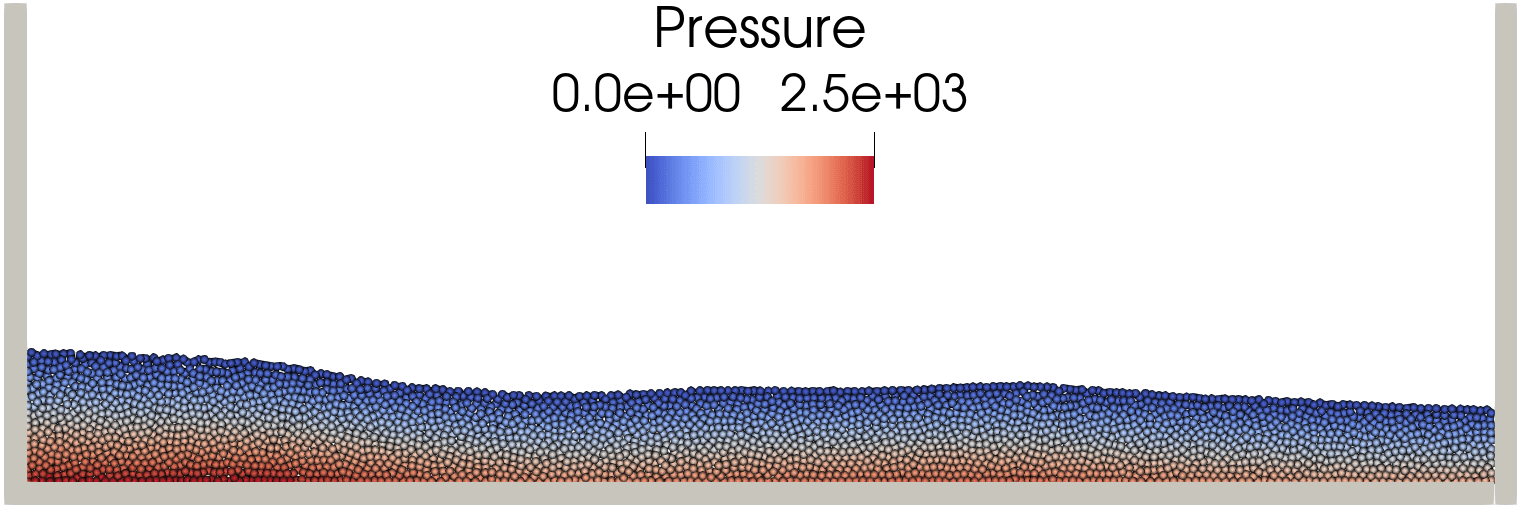}
            % \caption{Density variation of the \ac{wcsph}}
            \label{fig:WCSPH_dam_break_evolution_pressure_t=20}
        \end{subfigure}
        \vspace*{-8mm}
        \subcaption{\ac{ucsph}}
    \end{subfigure}
    
    \begin{subfigure}[t]{1.\linewidth}
        \begin{subfigure}{.495\linewidth}
            \centering
            \includegraphics[width=1\columnwidth]{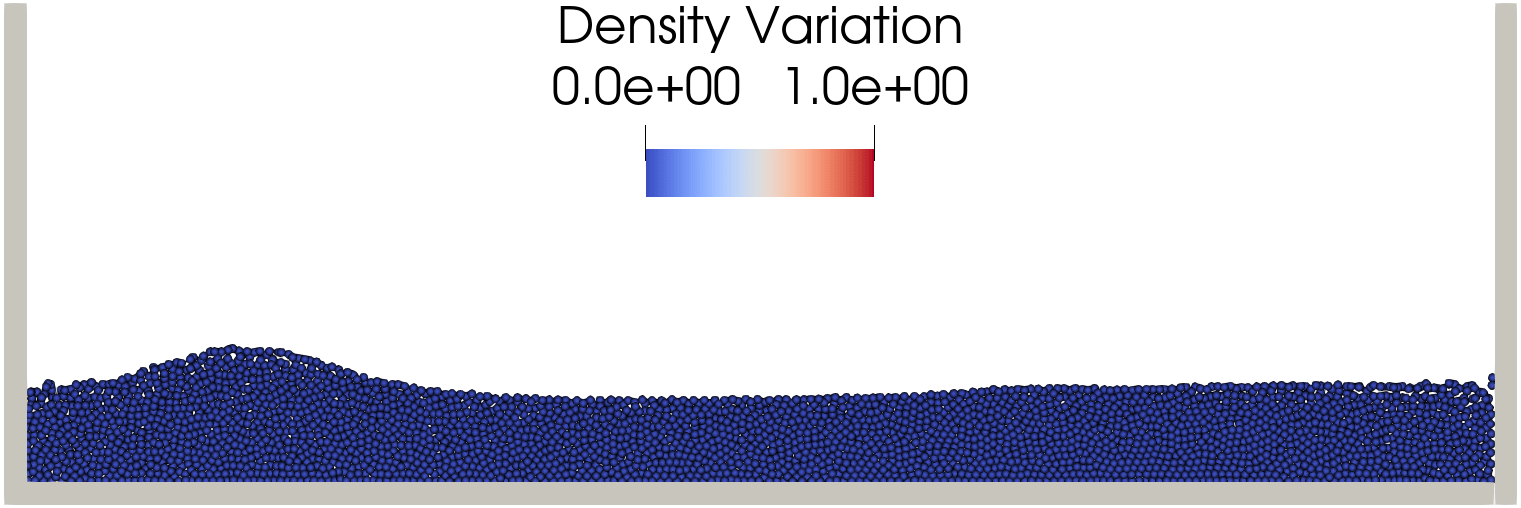}
            % \caption{Pressure of the \ac{ucsph}}
            \label{fig:UCSPH_dam_break_evolution_density_variation_t=20}
        \end{subfigure}
        \begin{subfigure}{.495\linewidth}
            \centering
            \includegraphics[width=1\columnwidth]{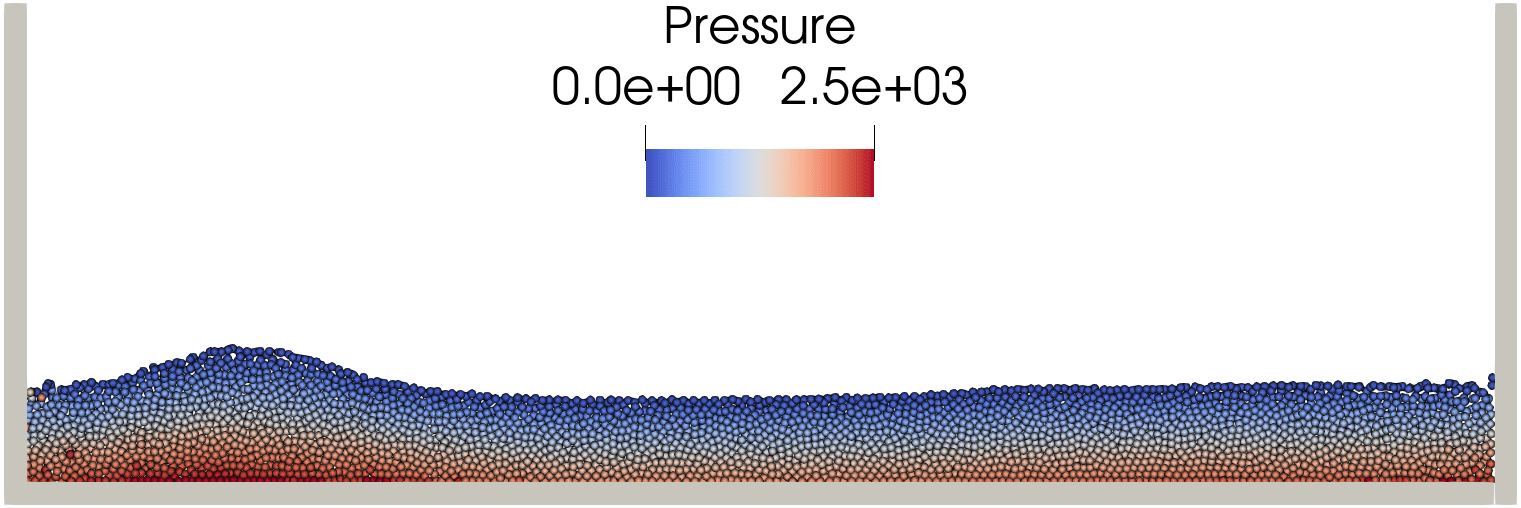}
            % \caption{Pressure of the \ac{wcsph}}
            \label{fig:WCSPH_dam_break_evolution_density_variation_t=20}
        \end{subfigure}
        \vspace*{-8mm}
        \subcaption{\ac{wcsph}}
    \end{subfigure}
    
    \caption{The density variation and pressure fields at T = 80.87 for the \ac{ucsph} and the \ac{wcsph} method.}
    \label{fig:dam_break_resting_fluid}

\end{figure}
Fig. \ref{fig:dam_break_resting_fluid} shows the density variation and the pressure field of both solvers at $T = 80.87$ ($t=20$ s). 
The pressure fields of both solutions are similar, but the density variations differ significantly. 
While the maximal density variation of the \ac{ucsph} method is close to $1\%$, the maximal variation from the \ac{wcsph} is in the order of $0.01\%$. 
Concurrently, the time step for \ac{ucsph} can be much larger compared to \ac{wcsph} due to the different speed of sounds.
The time steps with the \ac{ucsph} method are eight times higher than the one from the \ac{wcsph} method. 
For the entire simulation we observed a speed-up of factor $3.2$ for \ac{ucsph} compared to the \ac{wcsph} solver.

% INCLUDE THE SECTION CONCLUSION
\section{Conclusion}
\label{sec:conclusion}
In this work, we have discussed the weaknesses of using a constant speed of sound for the classical \ac{wcsph} method. 
In order to overcome these weaknesses, the Uniform Compressible SPH (UCSPH) method was introduced. 
The novel method relies on an algorithm that adjusts the speed of sound dependent on the present flow field. 
This results in a method with quasi constant maximal compressibility, even for highly changing flow fields. 
The variable speed of sound significantly influences the possible time step and, therefore, enables large speed-ups compared to the classical \ac{wcsph} method. 
In some cases, the variable speed of sound even positively affects the method's accuracy.
With three test cases (Taylor-Green vortex, falling droplet, and dam break), the method's robustness, accuracy, and performance are proven. 

%%       \citet{<label>} ==> Jones et al. [21]
%%       \citep{<label>} ==> [21]
%%

% \appendix

% \section{Sample Appendix Section}
% \label{sec:sample:appendix}
% Lorem ipsum dolor sit amet, consectetur adipiscing elit, sed do eiusmod tempor section incididunt ut labore et dolore magna aliqua. Ut enim ad minim veniam, quis nostrud exercitation ullamco laboris nisi ut aliquip ex ea commodo consequat. Duis aute irure dolor in reprehenderit in voluptate velit esse cillum dolore eu fugiat nulla pariatur. Excepteur sint occaecat cupidatat non proident, sunt in culpa qui officia deserunt mollit anim id est laborum.

\bibliographystyle{elsarticle-num-names}
\bibliography{main}

\begin{thebibliography}{35}
\expandafter\ifx\csname natexlab\endcsname\relax\def\natexlab#1{#1}\fi
\providecommand{\url}[1]{\texttt{#1}}
\providecommand{\href}[2]{#2}
\providecommand{\path}[1]{#1}
\providecommand{\DOIprefix}{doi:}
\providecommand{\ArXivprefix}{arXiv:}
\providecommand{\URLprefix}{URL: }
\providecommand{\Pubmedprefix}{pmid:}
\providecommand{\doi}[1]{\href{http://dx.doi.org/#1}{\path{#1}}}
\providecommand{\Pubmed}[1]{\href{pmid:#1}{\path{#1}}}
\providecommand{\bibinfo}[2]{#2}
\ifx\xfnm\relax \def\xfnm[#1]{\unskip,\space#1}\fi
%Type = Article
\bibitem[{Gingold and Monaghan(1977)}]{Gingold1977}
\bibinfo{author}{R.~A. Gingold}, \bibinfo{author}{J.~J. Monaghan},
\newblock \bibinfo{title}{Smoothed particle hydrodynamics: theory and application to non-spherical stars},
\newblock \bibinfo{journal}{Monthly Notices of the Royal Astronomical Society} \bibinfo{volume}{181} (\bibinfo{year}{1977}) \bibinfo{pages}{375--389}. \DOIprefix\doi{10.1093/mnras/181.3.375}.
%Type = Article
\bibitem[{Lucy(1977)}]{Lucy1977}
\bibinfo{author}{L.~B. Lucy},
\newblock \bibinfo{title}{A numerical approach to the testing of the fission hypothesis},
\newblock \bibinfo{journal}{The Astronomical Journal} \bibinfo{volume}{82} (\bibinfo{year}{1977}). \DOIprefix\doi{10.1086/112164}.
%Type = Article
\bibitem[{Monaghan and Kocharyan(1994)}]{Monaghan1994MultiPhase}
\bibinfo{author}{J.~J. Monaghan}, \bibinfo{author}{A.~Kocharyan},
\newblock \bibinfo{title}{Sph simulation of multi-phase flow},
\newblock \bibinfo{journal}{Computer Physics Communications} \bibinfo{volume}{87} (\bibinfo{year}{1994}) \bibinfo{pages}{225--235}. \DOIprefix\doi{https://doi.org/10.1016/0010-4655(94)00174-Z}.
%Type = Article
\bibitem[{Colagrossi and Landrini(2003)}]{Colagrossi2003}
\bibinfo{author}{A.~Colagrossi}, \bibinfo{author}{M.~Landrini},
\newblock \bibinfo{title}{Numerical simulation of interfacial flows by smoothed particle hydrodynamics},
\newblock \bibinfo{journal}{Journal of Computational Physics} \bibinfo{volume}{191} (\bibinfo{year}{2003}) \bibinfo{pages}{448--475}. \DOIprefix\doi{10.1016/s0021-9991(03)00324-3}.
%Type = Article
\bibitem[{Chen et~al.(2015)Chen, Zong, Liu, Zou, Li, and Shu}]{Chen2015}
\bibinfo{author}{Z.~Chen}, \bibinfo{author}{Z.~Zong}, \bibinfo{author}{M.~B. Liu}, \bibinfo{author}{L.~Zou}, \bibinfo{author}{H.~T. Li}, \bibinfo{author}{C.~Shu},
\newblock \bibinfo{title}{An sph model for multiphase flows with complex interfaces and large density differences},
\newblock \bibinfo{journal}{Journal of Computational Physics} \bibinfo{volume}{283} (\bibinfo{year}{2015}) \bibinfo{pages}{169--188}. \DOIprefix\doi{10.1016/j.jcp.2014.11.037}.
%Type = Article
\bibitem[{Monaghan(1994)}]{Monaghan1994FreeSurface}
\bibinfo{author}{J.~J. Monaghan},
\newblock \bibinfo{title}{Simulating free surface flows with sph},
\newblock \bibinfo{journal}{Journal of Computational Physics} \bibinfo{volume}{110} (\bibinfo{year}{1994}) \bibinfo{pages}{399--406}. \DOIprefix\doi{https://doi.org/10.1006/jcph.1994.1034}.
%Type = Article
\bibitem[{Violeau and Rogers(2016)}]{Violeau2016}
\bibinfo{author}{D.~Violeau}, \bibinfo{author}{B.~D. Rogers},
\newblock \bibinfo{title}{Smoothed particle hydrodynamics (sph) for free-surface flows: past, present and future},
\newblock \bibinfo{journal}{Journal of Hydraulic Research} \bibinfo{volume}{54} (\bibinfo{year}{2016}) \bibinfo{pages}{1--26}. \DOIprefix\doi{10.1080/00221686.2015.1119209}.
%Type = Article
\bibitem[{Holmes et~al.(2011)Holmes, Williams, and Tilke}]{Holmes2011}
\bibinfo{author}{D.~W. Holmes}, \bibinfo{author}{J.~R. Williams}, \bibinfo{author}{P.~Tilke},
\newblock \bibinfo{title}{Smooth particle hydrodynamics simulations of low reynolds number flows through porous media},
\newblock \bibinfo{journal}{International Journal for Numerical and Analytical Methods in Geomechanics} \bibinfo{volume}{35} (\bibinfo{year}{2011}) \bibinfo{pages}{419--437}. \DOIprefix\doi{10.1002/nag.898}.
%Type = Article
\bibitem[{Holmes and Pivonka(2021)}]{Holmes2021}
\bibinfo{author}{D.~W. Holmes}, \bibinfo{author}{P.~Pivonka},
\newblock \bibinfo{title}{Novel pressure inlet and outlet boundary conditions for smoothed particle hydrodynamics, applied to real problems in porous media flow},
\newblock \bibinfo{journal}{Journal of Computational Physics} \bibinfo{volume}{429} (\bibinfo{year}{2021}). \DOIprefix\doi{10.1016/j.jcp.2020.110029}.
%Type = Article
\bibitem[{Osorno et~al.(2021)Osorno, Schirwon, Kijanski, Sivanesapillai, Steeb, and Göddeke}]{Osorno2021}
\bibinfo{author}{M.~Osorno}, \bibinfo{author}{M.~Schirwon}, \bibinfo{author}{N.~Kijanski}, \bibinfo{author}{R.~Sivanesapillai}, \bibinfo{author}{H.~Steeb}, \bibinfo{author}{D.~Göddeke},
\newblock \bibinfo{title}{A cross-platform, high-performance sph toolkit for image-based flow simulations on the pore scale of porous media},
\newblock \bibinfo{journal}{Computer Physics Communications} \bibinfo{volume}{267} (\bibinfo{year}{2021}). \DOIprefix\doi{10.1016/j.cpc.2021.108059}.
%Type = Article
\bibitem[{Morris(2000)}]{Morris2000}
\bibinfo{author}{J.~P. Morris},
\newblock \bibinfo{title}{Simulating surface tension with smoothed particle hydrodynamics},
\newblock \bibinfo{journal}{International Journal for Numerical Methods in Fluids}  (\bibinfo{year}{2000}). \DOIprefix\doi{https://doi.org/10.1002/1097-0363(20000615)33:3\%3C333::AID-FLD11\%3E3.0.CO;2-7}.
%Type = Article
\bibitem[{Hu and Adams(2006)}]{Hu2006}
\bibinfo{author}{X.~Y. Hu}, \bibinfo{author}{N.~A. Adams},
\newblock \bibinfo{title}{A multi-phase sph method for macroscopic and mesoscopic flows},
\newblock \bibinfo{journal}{Journal of Computational Physics} \bibinfo{volume}{213} (\bibinfo{year}{2006}) \bibinfo{pages}{844--861}. \DOIprefix\doi{10.1016/j.jcp.2005.09.001}.
%Type = Article
\bibitem[{Blank et~al.(2023)Blank, Nair, and Pöschel}]{Blank2023}
\bibinfo{author}{M.~Blank}, \bibinfo{author}{P.~Nair}, \bibinfo{author}{T.~Pöschel},
\newblock \bibinfo{title}{Modeling surface tension in smoothed particle hydrodynamics using young–laplace pressure boundary condition},
\newblock \bibinfo{journal}{Computer Methods in Applied Mechanics and Engineering} \bibinfo{volume}{406} (\bibinfo{year}{2023}). \DOIprefix\doi{10.1016/j.cma.2023.115907}.
%Type = Article
\bibitem[{Adami et~al.(2010)Adami, Hu, and Adams}]{Adami2010Surfactant}
\bibinfo{author}{S.~Adami}, \bibinfo{author}{X.~Y. Hu}, \bibinfo{author}{N.~A. Adams},
\newblock \bibinfo{title}{A conservative sph method for surfactant dynamics},
\newblock \bibinfo{journal}{Journal of Computational Physics} \bibinfo{volume}{229} (\bibinfo{year}{2010}) \bibinfo{pages}{1909--1926}. \DOIprefix\doi{10.1016/j.jcp.2009.11.015}.
%Type = Article
\bibitem[{Kim et~al.(2023)Kim, Yoon, Yoo, and Kim}]{Kim2023}
\bibinfo{author}{J.-W. Kim}, \bibinfo{author}{S.-J. Yoon}, \bibinfo{author}{T.-S. Yoo}, \bibinfo{author}{E.~S. Kim},
\newblock \bibinfo{title}{Modelling and analysis of salt-convection effect on oxide reduction process for uranium oxides using smoothed particle hydrodynamics},
\newblock \bibinfo{journal}{International Journal of Heat and Mass Transfer} \bibinfo{volume}{206} (\bibinfo{year}{2023}). \DOIprefix\doi{10.1016/j.ijheatmasstransfer.2023.123965}.
%Type = Article
\bibitem[{Cummins and Rudman(1999)}]{Cummins1999}
\bibinfo{author}{S.~J. Cummins}, \bibinfo{author}{M.~Rudman},
\newblock \bibinfo{title}{An sph projection method},
\newblock \bibinfo{journal}{Journal of Computational Physics} \bibinfo{volume}{152} (\bibinfo{year}{1999}) \bibinfo{pages}{584--607}. \DOIprefix\doi{10.1006/jcph.1999.6246}.
%Type = Article
\bibitem[{Lind et~al.(2012)Lind, Xu, Stansby, and Rogers}]{Lind2012}
\bibinfo{author}{S.~J. Lind}, \bibinfo{author}{R.~Xu}, \bibinfo{author}{P.~K. Stansby}, \bibinfo{author}{B.~D. Rogers},
\newblock \bibinfo{title}{Incompressible smoothed particle hydrodynamics for free-surface flows: A generalised diffusion-based algorithm for stability and validations for impulsive flows and propagating waves},
\newblock \bibinfo{journal}{Journal of Computational Physics} \bibinfo{volume}{231} (\bibinfo{year}{2012}) \bibinfo{pages}{1499--1523}. \DOIprefix\doi{10.1016/j.jcp.2011.10.027}.
%Type = Article
\bibitem[{Solenthaler and Pajarola(2009)}]{Solenthaler2009}
\bibinfo{author}{B.~Solenthaler}, \bibinfo{author}{R.~Pajarola},
\newblock \bibinfo{title}{Predictive-corrective incompressible sph},
\newblock \bibinfo{journal}{ACM Transactions on Graphics} \bibinfo{volume}{28} (\bibinfo{year}{2009}) \bibinfo{pages}{1--6}. \DOIprefix\doi{10.1145/1531326.1531346}.
%Type = Article
\bibitem[{Ihmsen et~al.(2014)Ihmsen, Cornelis, Solenthaler, Horvath, and Teschner}]{Ihmsen2014}
\bibinfo{author}{M.~Ihmsen}, \bibinfo{author}{J.~Cornelis}, \bibinfo{author}{B.~Solenthaler}, \bibinfo{author}{C.~Horvath}, \bibinfo{author}{M.~Teschner},
\newblock \bibinfo{title}{Implicit incompressible sph},
\newblock \bibinfo{journal}{IEEE Trans Vis Comput Graph} \bibinfo{volume}{20} (\bibinfo{year}{2014}) \bibinfo{pages}{426--35}. \URLprefix \url{https://www.ncbi.nlm.nih.gov/pubmed/24434223}. \DOIprefix\doi{10.1109/TVCG.2013.105}.
%Type = Inproceedings
\bibitem[{Bender and Koschier(2015)}]{Bender2015}
\bibinfo{author}{J.~Bender}, \bibinfo{author}{D.~Koschier},
\newblock \bibinfo{title}{Divergence-free smoothed particle hydrodynamics},
\newblock in: \bibinfo{booktitle}{Proceedings of the 14th ACM SIGGRAPH / Eurographics Symposium on Computer Animation}, \bibinfo{year}{2015}, p. \bibinfo{pages}{147–155}. \URLprefix \url{https://doi.org/10.1145/2786784.2786796}. \DOIprefix\doi{10.1145/2786784.2786796}.
%Type = Article
\bibitem[{Peng et~al.(2019)Peng, Bauinger, Szewc, Wu, and Cao}]{Peng2019}
\bibinfo{author}{C.~Peng}, \bibinfo{author}{C.~Bauinger}, \bibinfo{author}{K.~Szewc}, \bibinfo{author}{W.~Wu}, \bibinfo{author}{H.~Cao},
\newblock \bibinfo{title}{An improved predictive-corrective incompressible smoothed particle hydrodynamics method for fluid flow modelling},
\newblock \bibinfo{journal}{Journal of Hydrodynamics} \bibinfo{volume}{31} (\bibinfo{year}{2019}) \bibinfo{pages}{654--668}. \DOIprefix\doi{10.1007/s42241-019-0058-5}.
%Type = Article
\bibitem[{Chen et~al.(2013)Chen, Zong, Liu, and Li}]{Chen2013}
\bibinfo{author}{Z.~Chen}, \bibinfo{author}{Z.~Zong}, \bibinfo{author}{M.~B. Liu}, \bibinfo{author}{H.~T. Li},
\newblock \bibinfo{title}{A comparative study of truly incompressible and weakly compressible sph methods for free surface incompressible flows},
\newblock \bibinfo{journal}{International Journal for Numerical Methods in Fluids}  (\bibinfo{year}{2013}) \bibinfo{pages}{n/a--n/a}. \DOIprefix\doi{10.1002/fld.3824}.
%Type = Article
\bibitem[{Lee et~al.(2008)Lee, Moulinec, Xu, Violeau, Laurence, and Stansby}]{Lee2008}
\bibinfo{author}{E.~S. Lee}, \bibinfo{author}{C.~Moulinec}, \bibinfo{author}{R.~Xu}, \bibinfo{author}{D.~Violeau}, \bibinfo{author}{D.~Laurence}, \bibinfo{author}{P.~Stansby},
\newblock \bibinfo{title}{Comparisons of weakly compressible and truly incompressible algorithms for the sph mesh free particle method},
\newblock \bibinfo{journal}{Journal of Computational Physics} \bibinfo{volume}{227} (\bibinfo{year}{2008}) \bibinfo{pages}{8417--8436}. \DOIprefix\doi{10.1016/j.jcp.2008.06.005}.
%Type = Book
\bibitem[{Cole(1948)}]{Cole1948}
\bibinfo{author}{R.~H. Cole}, \bibinfo{title}{Underwater explosions.}, \bibinfo{publisher}{Princeton Univ. Press}, \bibinfo{year}{1948}.
%Type = Article
\bibitem[{Morris et~al.(1997)Morris, Fox, and Zhu}]{Morris1997}
\bibinfo{author}{J.~P. Morris}, \bibinfo{author}{P.~J. Fox}, \bibinfo{author}{Y.~Zhu},
\newblock \bibinfo{title}{Modeling low reynolds number incompressible flows using sph},
\newblock \bibinfo{journal}{Journal of Computational Physics} \bibinfo{volume}{136} (\bibinfo{year}{1997}) \bibinfo{pages}{214--226}. \URLprefix \url{https://www.sciencedirect.com/science/article/pii/S0021999197957764}. \DOIprefix\doi{10.1006/jcph.1997.5776}.
%Type = Article
\bibitem[{Monaghan and Kos(1999)}]{Monaghan1999}
\bibinfo{author}{J.~J. Monaghan}, \bibinfo{author}{A.~Kos},
\newblock \bibinfo{title}{Solitary waves on a cretan beach},
\newblock \bibinfo{journal}{Journal of Waterway, Port, Coastal, and Ocean Engineering} \bibinfo{volume}{125} (\bibinfo{year}{1999}) \bibinfo{pages}{145--155}. \DOIprefix\doi{10.1061/(asce)0733-950x(1999)125:3(145)}.
%Type = Article
\bibitem[{Crespo et~al.(2015)Crespo, Domínguez, Rogers, Gómez-Gesteira, Longshaw, Canelas, Vacondio, Barreiro, and García-Feal}]{Crespo2015}
\bibinfo{author}{A.~J.~C. Crespo}, \bibinfo{author}{J.~M. Domínguez}, \bibinfo{author}{B.~D. Rogers}, \bibinfo{author}{M.~Gómez-Gesteira}, \bibinfo{author}{S.~Longshaw}, \bibinfo{author}{R.~Canelas}, \bibinfo{author}{R.~Vacondio}, \bibinfo{author}{A.~Barreiro}, \bibinfo{author}{O.~García-Feal},
\newblock \bibinfo{title}{Dualsphysics: Open-source parallel cfd solver based on smoothed particle hydrodynamics (sph)},
\newblock \bibinfo{journal}{Computer Physics Communications} \bibinfo{volume}{187} (\bibinfo{year}{2015}) \bibinfo{pages}{204--216}. \DOIprefix\doi{10.1016/j.cpc.2014.10.004}.
%Type = Article
\bibitem[{Hahn et~al.(2018)Hahn, Adami, and Förstner}]{Hahn2018}
\bibinfo{author}{M.~Hahn}, \bibinfo{author}{S.~Adami}, \bibinfo{author}{R.~Förstner},
\newblock \bibinfo{title}{Computational modeling of nonlinear propellant sloshing for spacecraft aocs applications},
\newblock \bibinfo{journal}{CEAS Space Journal} \bibinfo{volume}{10} (\bibinfo{year}{2018}) \bibinfo{pages}{441--451}. \DOIprefix\doi{10.1007/s12567-018-0216-6}.
%Type = Article
\bibitem[{Zhang et~al.(2023)Zhang, Zhang, Zhang, and Hu}]{Zhang2022}
\bibinfo{author}{S.~Zhang}, \bibinfo{author}{W.~Zhang}, \bibinfo{author}{C.~Zhang}, \bibinfo{author}{X.~Hu},
\newblock \bibinfo{title}{A lagrangian free-stream boundary condition for weakly compressible smoothed particle hydrodynamics},
\newblock \bibinfo{journal}{Journal of Computational Physics} \bibinfo{volume}{490} (\bibinfo{year}{2023}) \bibinfo{pages}{112303}. \URLprefix \url{https://www.sciencedirect.com/science/article/pii/S0021999123003984}. \DOIprefix\doi{https://doi.org/10.1016/j.jcp.2023.112303}.
%Type = Article
\bibitem[{Verlet(1967)}]{Verlet1967}
\bibinfo{author}{L.~Verlet},
\newblock \bibinfo{title}{Computer "experiments" on classical fluids. i. thermodynamical properties of lennard-jones molecules},
\newblock \bibinfo{journal}{Physical Review} \bibinfo{volume}{159} (\bibinfo{year}{1967}). \URLprefix \url{https://journals.aps.org/pr/pdf/10.1103/PhysRev.159.98}. \DOIprefix\doi{10.1103/PhysRev.159.98}.
%Type = Article
\bibitem[{Ramachandran and Puri(2019)}]{Ramachandran2019}
\bibinfo{author}{P.~Ramachandran}, \bibinfo{author}{K.~Puri},
\newblock \bibinfo{title}{Entropically damped artificial compressibility for sph},
\newblock \bibinfo{journal}{Computers \& Fluids} \bibinfo{volume}{179} (\bibinfo{year}{2019}) \bibinfo{pages}{579--594}. \DOIprefix\doi{10.1016/j.compfluid.2018.11.023}.
%Type = Article
\bibitem[{Adami et~al.(2013)Adami, Hu, and Adams}]{Adami2013}
\bibinfo{author}{S.~Adami}, \bibinfo{author}{X.~Y. Hu}, \bibinfo{author}{N.~A. Adams},
\newblock \bibinfo{title}{A transport-velocity formulation for smoothed particle hydrodynamics},
\newblock \bibinfo{journal}{Journal of Computational Physics} \bibinfo{volume}{241} (\bibinfo{year}{2013}) \bibinfo{pages}{292--307}. \DOIprefix\doi{10.1016/j.jcp.2013.01.043}.
%Type = Phdthesis
\bibitem[{Buchner(2002)}]{Buchner2002}
\bibinfo{author}{B.~Buchner}, \bibinfo{title}{Green Water on Ship-type Offshore Structures}, \bibinfo{type}{Thesis}, Delft University of Technology, \bibinfo{year}{2002}.
%Type = Article
\bibitem[{Adami et~al.(2012)Adami, Hu, and Adams}]{Adami2012}
\bibinfo{author}{S.~Adami}, \bibinfo{author}{X.~Y. Hu}, \bibinfo{author}{N.~A. Adams},
\newblock \bibinfo{title}{A generalized wall boundary condition for smoothed particle hydrodynamics},
\newblock \bibinfo{journal}{Journal of Computational Physics} \bibinfo{volume}{231} (\bibinfo{year}{2012}) \bibinfo{pages}{7057--7075}. \DOIprefix\doi{10.1016/j.jcp.2012.05.005}.
%Type = Article
\bibitem[{Marrone et~al.(2011)Marrone, Antuono, Colagrossi, Colicchio, Le~Touzé, and Graziani}]{Marrone2011}
\bibinfo{author}{S.~Marrone}, \bibinfo{author}{M.~Antuono}, \bibinfo{author}{A.~Colagrossi}, \bibinfo{author}{G.~Colicchio}, \bibinfo{author}{D.~Le~Touzé}, \bibinfo{author}{G.~Graziani},
\newblock \bibinfo{title}{$\delta$-sph model for simulating violent impact flows},
\newblock \bibinfo{journal}{Computer Methods in Applied Mechanics and Engineering} \bibinfo{volume}{200} (\bibinfo{year}{2011}) \bibinfo{pages}{1526--1542}. \DOIprefix\doi{10.1016/j.cma.2010.12.016}.

\end{thebibliography}





% \end{thebibliography}
\end{document}